\documentclass[conference]{IEEEtran}
\usepackage{cite}
\usepackage{amsmath,amssymb,amsfonts}
\usepackage{algorithmic}
\usepackage{graphicx}
\usepackage{textcomp}
\usepackage{xcolor}
\usepackage[hyphens]{url}
\usepackage{siunitx}
\def\BibTeX{{\rm B\kern-.05em{\sc i\kern-.025em b}\kern-.08em
    T\kern-.1667em\lower.7ex\hbox{E}\kern-.125emX}}

\title{MAC-DO: An Efficient Output-Stationary GEMM Accelerator for CNNs Using DRAM Technology}

\author{ MINKI JEONG \hspace{1em}   Wanyeong Jung \\ \vspace{-1em} School of Electrical Engineering \\ \vspace{-1em} \\ \\ Korea Advanced Institute of Science and Technology (KAIST) \\ \{mkaistk, wanyeong\}@kaist.ac.kr }

\begin{document}
\maketitle
\thispagestyle{plain}
\pagestyle{plain}


\begin{abstract}

DRAM-based in-situ accelerators have shown their potential in addressing the memory wall challenge of the traditional von Neumann architecture. Such accelerators exploit charge sharing or logic circuits for simple logic operations at the DRAM subarray level. However, their throughput is limited due to low array utilization, as only a few row cells in a DRAM array participate in operations while most rows remain deactivated. Moreover, they require many cycles for more complex operations such as a multi-bit multiply-accumulate (MAC) operation, resulting in significant data access and movement and potentially worsening power efficiency.

To overcome these limitations, this paper presents MAC-DO, an efficient and low-power DRAM-based in-situ accelerator. Compared to previous DRAM-based in-situ accelerators, a MAC-DO cell, consisting of two 1T1C DRAM cells (two transistors and two capacitors), innately supports a multi-bit MAC operation within a single cycle, ensuring good linearity and compatibility with existing 1T1C DRAM cells and array structures. This achievement is facilitated by a novel analog computation method utilizing charge steering. Additionally, MAC-DO enables concurrent individual MAC operations in each MAC-DO cell without idle cells, significantly improving throughput and energy efficiency. As a result, a MAC-DO array efficiently can accelerate matrix multiplications based on output stationary mapping, supporting the majority of computations performed in deep neural networks (DNNs). Furthermore, a MAC-DO array efficiently reuses three types of data (input, weight and output), minimizing data movement.

Our evaluation using transistor-level simulation shows that a test MAC-DO array with 16$\times$16 MAC-DO cells achieves 120.96 TOPS/W. Furthermore, our system level evaluation demonstrates that the MAC-DO array marks $>$300$\times$ data movement reduction and $>$17.9$\times$ speedup for CNNs over previous DRAM-based in-situ accelerators.
\end{abstract}

\section{Introduction}
The demand for efficient processing of large data volumes is increasing, particularly in edge devices, which are often constrained by battery limitations \cite{shi2016edge,satyanarayanan2017emergence}.
However, traditional von Neumann architecture faces inefficiencies when dealing with extensive data due to costly data movement and a bottleneck between the CPU and memory \cite{han2016eie,villa2014scaling}.
To address these challenges, GPUs have been employed to mitigate the bottleneck problem by performing parallel operations and reusing data multiple times after reading the data from memory. Despite these advantages, GPUs still present certain limitations, including high cost and energy consumption \cite{collange2009power}.

\begin{figure}[t]
	\centering
	\includegraphics[width=0.79\columnwidth]{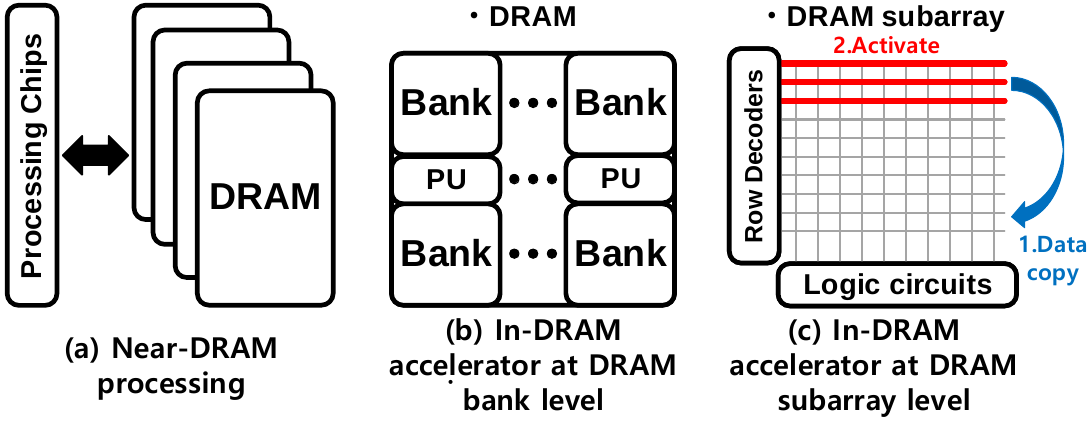}
	\caption{Different types of DRAM-based processing techniques} 
	\label{figure32}
\end{figure}

Accelerator-in-memory (AiM) architecture addresses the cost of data movement and memory bottleneck by enabling parallel computations within the memory itself \cite{ahn2015scalable,xie2021spacea,bojnordi2016memristive,kim2018image,gu2020ipim,imani2019floatpim,srivastava2018promise}. One of the key design considerations for AiM architecture is how efficiently and effectively it can process a significant amount of data in parallel while operating within restricted memory area constraints. In this context, DRAM-based Accelerator-in-Memory (AiM) architectures offer a promising solution due to their larger memory capacity compared to other memory technologies like SRAM \cite{biswas2018conv,dong202015}. The vertical stacking of just 1 transistor and 1 capacitor in a DRAM cell results in higher cell density compared to other memory cells, enabling efficient space utilization. Consequently, maximizing the utilization of DRAM cells becomes an advantageous strategy when dealing with area constraints while processing substantial amounts of data. Moreover, they offer the advantage of being located near the main memory, typically DRAM in computing systems, thereby minimizing the distance and amount of data movement between the processor and the main memory. 
Actually, many near-DRAM-based processing techniques and architectures (Figure \ref{figure32}-(a)) have been developed and adopted in industry such as Google TPUs (HBM memories are used since TPUv2 \cite{TPUv2v39351692}). However, most near-DRAM-based approaches still require frequent off-chip data movement between individual DRAMs and data processing chips, resulting in higher costs compared to on-chip data movement \cite{horowitz20141}.

To minimize off-chip data movement, recent advancements in in-DRAM data processing have introduced two types of in-DRAM accelerators: those operating at the DRAM bank level \cite{shin2018mcdram,lee20221ynm}, and those processing data at the subarray level \cite{li2017drisa,seshadri2017ambit,angizi2019redram,li2018scope,xin2020elp2im}. The former \cite{shin2018mcdram,lee20221ynm} places processing units (PUs) near DRAM banks (Figure \ref{figure32}-(b)) to enable efficient data processing through bank-level parallelism. However, their performance remains limited as they are unable to fully utilize the internal DRAM bandwidth, which is designed to match the device I/O bandwidth.
On the other hand, the latter \cite{li2017drisa,seshadri2017ambit,angizi2019redram,li2018scope,xin2020elp2im} leverages the full internal DRAM bandwidth by performing data processing at the subarray level (Figure \ref{figure32}-(c)). This approach, referred to as "DRAM-based in-situ accelerator" no longer relies on common memory access operations for system memory, and can be adapted for higher computational ability at the cost of more complicated array control.

However, previous DRAM-based in-situ accelerators face several challenges. They are restricted to performing simple logic operations like NOR and NOT within a computation cycle. For instance, \cite{angizi2019redram} incorporates digital logic gates outside the DRAM array for logic operations. ELP2IM \cite{xin2020elp2im} employs charge sharing on bit-lines (BL) to execute logic operations. While logic operations in DRAM-based accelerators can handle various tasks through repetitive bitwise operations, they suffer from significant data access and movement between DRAM cells and logic cells, which are typically incompatible in technology process \cite{kim1999assessing}. Additionally, prior to a logic operation, data must be copied from the stored location to other rows or DRAM arrays since the DRAM read process is destructive.
This inefficiency becomes particularly noticeable for more complex operations like multi-bit multiply-accumulate (MAC) operations, degrading the overall system performance. Even a single multi-bit multiplication can take several tens of cycles \cite{li2018scope,li2017drisa}, which reduces the power efficiency for MAC operations. 
Furthermore, these architectures suffer from limited data reusability and parallel computation because only one logic operation is performed on a bit-line per computation cycle. 
Perhaps most importantly, the throughput of these accelerators remains restricted because only a few rows participate in operations, leaving the majority of cells inactive during a computation cycle (see Figure \ref{figure32}-(c)). 
While these methods may suffice for enhancing data-intensive tasks where memory bandwidth is a bottleneck, they are not suitable for accelerating compute-intensive tasks, such as convolutions, which account for over 90\% of computations and runtime in CNN operations \cite{chen2016eyeriss} due to their low throughput and power efficiency.

\begin{figure}[t]
	\centering
	\includegraphics[width=0.8\columnwidth]{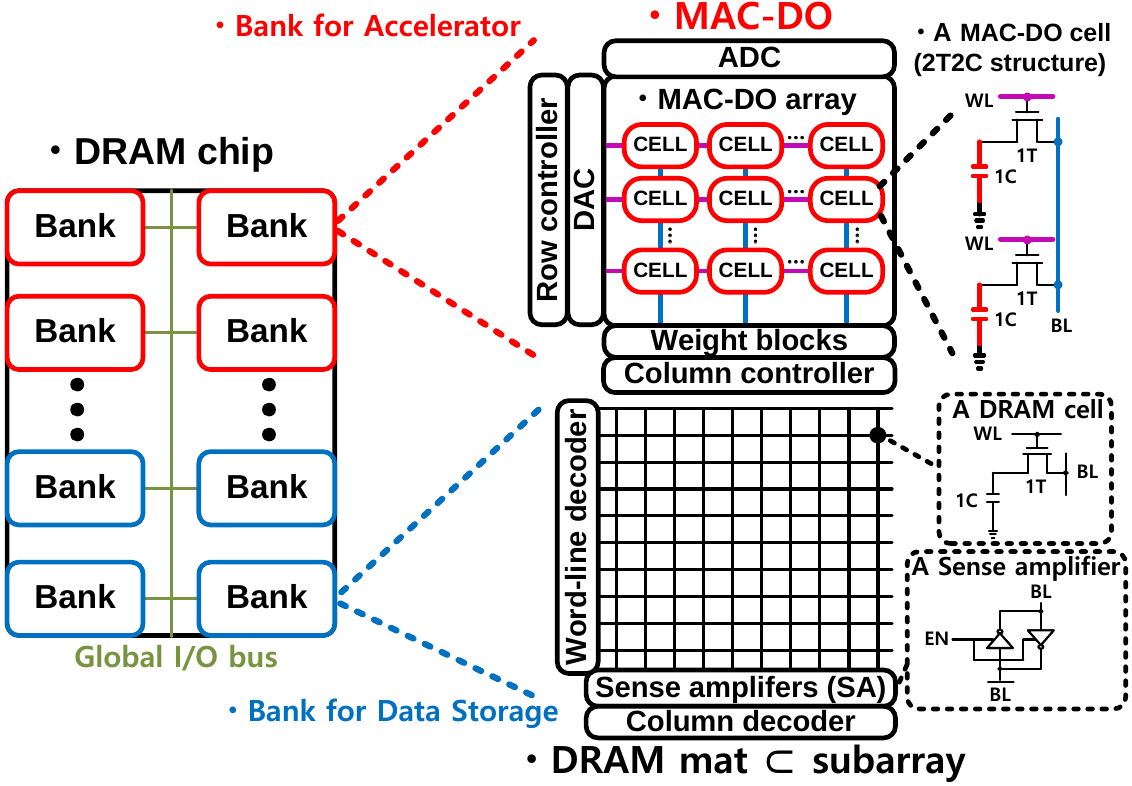}
	\caption{Proposed MAC-DO architecture} 
	\label{figure33}
\end{figure}

To address the challenges faced by previous DRAM-based in-situ accelerators, this paper presents MAC-DO, a high-performance DRAM-based in-situ accelerator designed to efficiently perform MAC operations within a DRAM array. Figure \ref{figure33} illustrates the proposed MAC-DO architecture, and Section \uppercase\expandafter{\romannumeral3} provides detailed information about each block of MAC-DO circuits. Unlike many prior DRAM-based accelerators, a MAC-DO cell, consisting of two 1T1C cells in a DRAM array, inherently supports a single-cycle MAC operation with multi-bit precision inputs and weights.  This achievement is made possible by adopting a novel analog computing mechanism based on charge steering \cite{razavi2013charge}, originally proposed for high-speed analog and mixed-signal circuits. Furthermore, a MAC-DO array efficiently accelerates matrix multiplications for convolutions using output stationary mapping \cite{moons201714,du2015shidiannao,reagen2021cheetah}, as all MAC-DO cells in a DRAM array can participate in individual MAC computations simultaneously without idle cells. This minimizes the overall data movement cost since each row and column of the MAC-DO cells share the same data, respectively, as many times as the array size. 
As a result, MAC-DO achieves significantly higher throughput and improved energy efficiency compared to previous DRAM-based in-situ accelerators, despite utilizing only a small portion of banks or mats in a DRAM chip to offset its area overhead.
Furthermore, MAC-DO's overall integration cost is expected to be low, while leveraging the advantages of analog MAC operations that offer higher energy efficiency at the expense of accuracy compared to digital methods. This extension enables the applicability of AI to low-cost and low-power edge devices \cite{venkataramani2021rapid}.

The main contributions of this paper are listed as follows:

\begin{itemize}

\item MAC-DO implements a charge-based analog MAC operation between multi-bit signed inputs and weights in a single cycle.
\item A MAC-DO cell exploits charge steering technique that was originally used for analog and mixed-signal applications only. This paper first proves its applicability in highly parallel analog computing by proposing a 2-D array architecture and control methods that are supported by extensive transistor-level simulation data. MAC-DO can take advantages of charge steering, including high-speed operation, good linearity and reliability, for analog computing as well.
\item MAC-DO is compatible with modern DRAM arrays directly without any modifications to existing DRAM cells. Hence, the overall integration cost of MAC-DO would be low. 
\item To the best of our knowledge, MAC-DO is the only architecture that can fully utilize the entire DRAM array for MAC operations. Every MAC-DO cell in a DRAM array participates in MAC computations simultaneously using output stationary mapping, generating different partial sums for efficient matrix multiplications in convolutions. Additionally, MAC-DO maximizes data reusability within the array, minimizing data movement costs. As a result, MAC-DO significantly improves throughput and energy efficiency, even with a small portion of banks or mats in DRAM.
\end{itemize}

The rest of this paper is organized as follows. Section \uppercase\expandafter{\romannumeral2} introduces the background of DRAM and outer product for matrix multiplications. Section \uppercase\expandafter{\romannumeral3} describes charge-steering topology for MAC-DO, actual MAC-DO circuits and its operations. Section \uppercase\expandafter{\romannumeral4} demonstrates digital and analog correction for negative weight and non-linear effects. Section \uppercase\expandafter{\romannumeral5} explains the evaluation methodology for performance test. Section \uppercase\expandafter{\romannumeral6} presents the evaluation results of MAC-DO. Section \uppercase\expandafter{\romannumeral7} concludes this paper.

\section{Background}

\subsection{DRAM and Its Operation}

A DRAM chip (Figure \ref{figure33}) consists of multiple banks connected by a global shared bus. Each bank is composed of subarray groups and each of them includes several cell matrices (mat), which is the basic unit of the DRAM chip. Each mat has its own independent sense amplifier (SA) row, a word-line (WL) decoder and a DRAM array. A SA amplifies the signal on a bit-line (BL) and quantizes it. A WL decoder controls WLs to write or read data. There are numerous DRAM cells inside a DRAM array and each cell consists of an access transistor and a cell capacitor, called a 1T1C cell. The access transistors in a row are activated to write data into each cell or to read the data stored in each cell through WLs and BLs. For example, when writing data '1' into a cell, the corresponding WL is activated and its cell capacitor is charged to a high voltage through the corresponding BL. On the other hand, if the cell capacitor is discharged, the cell stores data '0'. In a read process, a row of access transistors is turned on and the stored data are read through the SAs.

\subsection{Matrix Multiplications through Iterative Outer Products}
\begin{figure}[!h]
	\centering
	\includegraphics[width=0.9\columnwidth]{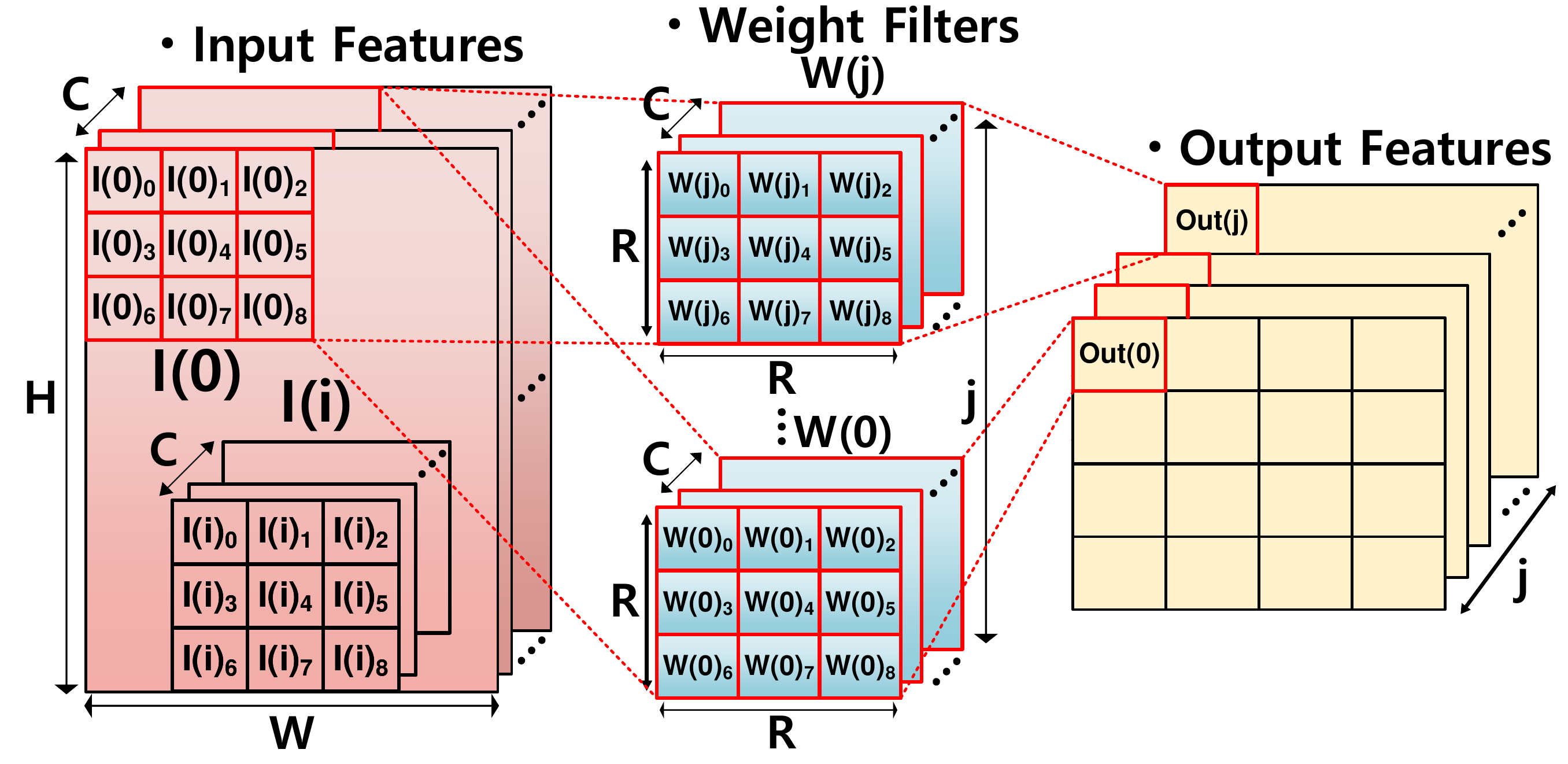}
	\caption{Convolutions}
	\label{figure30}
\end{figure}
\begin{figure}[!t]
	\centering
	\includegraphics[width=0.9\columnwidth]{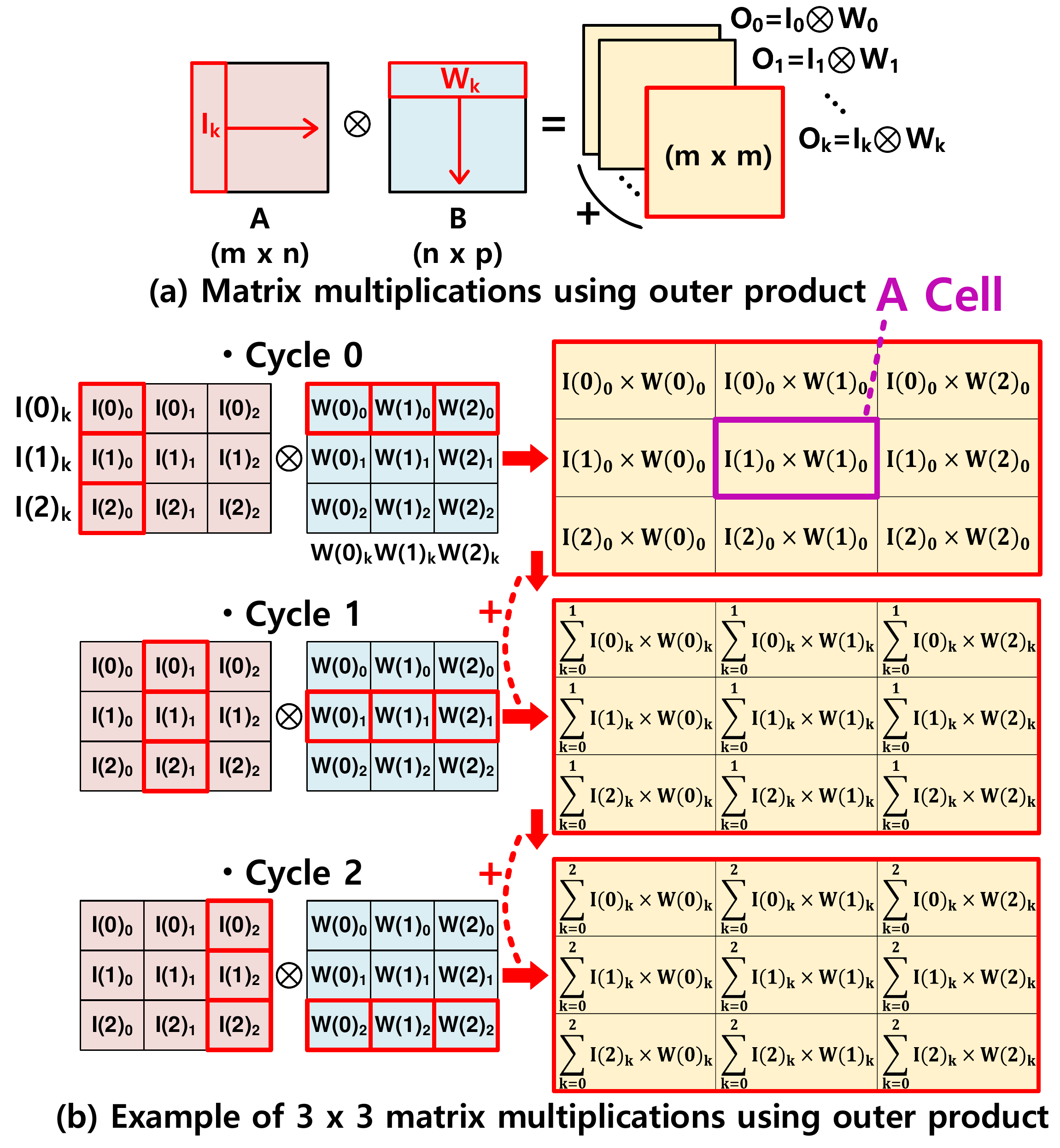}
	\caption{Matrix multiplications through iterative outer products}
	\label{figure27}
\end{figure}

A convolution layer (Figure \ref{figure30}) can be processed by repeating matrix multiplications \cite{vasudevan2017parallel,ofir2022smm}, and multiplication of two matrices $A$ and $B$ can be performed through iterative outer products between columns of $A$ and rows of $B$ \cite{pal2018outerspace}. First, each column of matrix $A$ (=$I_k$) is multiplied by the corresponding row of matrix $B$ (=$W_k$) as shown in Figure \ref{figure27}, forming a partial product (=matrix $O_k$). Then, the partial products are added up, resulting in the product of $A$ and $B$.
\begin{equation}
	\sum_{k}O_i = \sum_{k}I_k \times W_k = \begin{bmatrix} \sum_{k}A_{ik} \times B_{kj} \end{bmatrix} = A \times B
	\label{equation16}
\end{equation}

As this multiplication process involves a series of additions of partial product matrices with the same dimension, it can be easily mapped on a two-dimensional output-stationary MAC array. In each array cell, its horizontal input is multiplied by the vertical input, and the result keeps accumulated in the same cell into an output matrix element (="Output Stationary"). Figure \ref{figure27}-(b) shows an example of matrix multiplication between two 3$\times$3 matrices. Each cell performs individual MAC operations at each cycle and the final output matrix is formed after all cycles finish. 

Section \uppercase\expandafter{\romannumeral3}-C explains how a MAC-DO cell operates as a MAC unit, and Section \uppercase\expandafter{\romannumeral3}-E
shows how the MAC-DO cells can be combined into an output-stationary array that performs matrix multiplications for convolutions.

\section{MAC-DO and Its Operation}

\subsection{Charge-Steering Amplifier}

MAC-DO is based on a charge-steering topology\cite{razavi2013charge}. Charge steering is originally for a discrete-time analog amplifier, offering high-speed and low-power amplification compared with traditional current-steering amplifiers. As far as we know, its application in high-performance analog computing is first proposed in this paper.

\begin{figure}[t]
	\centering
	\includegraphics[width=0.75\columnwidth]{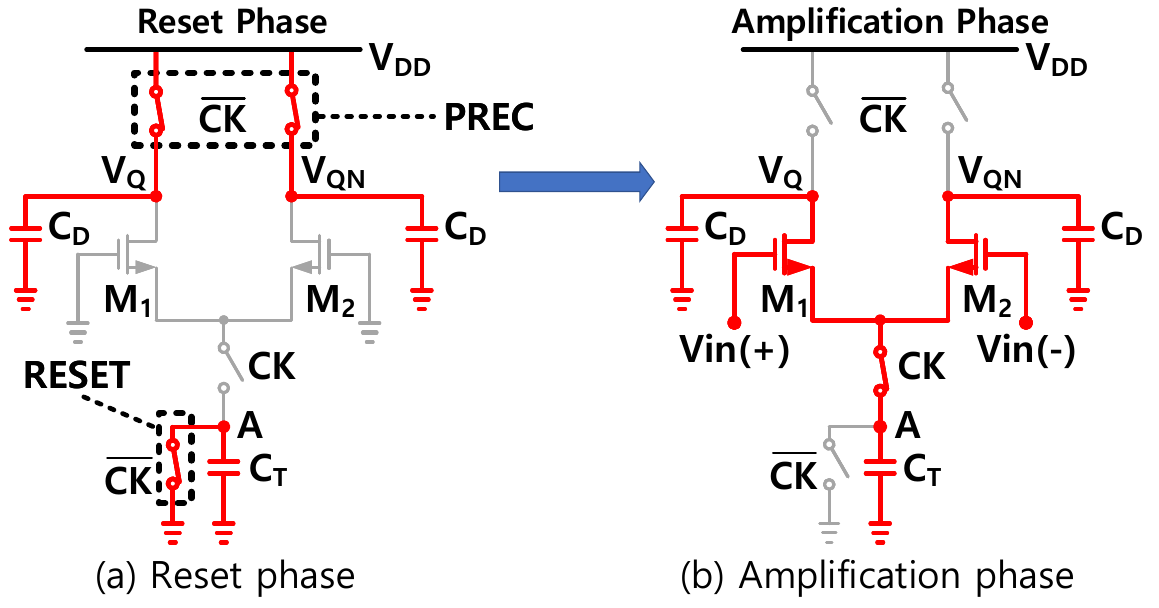}
	\caption{The operation of a charge-steering amplifier}
	\label{figure1}
\end{figure}

As shown in Figure \ref{figure1}, the charge-steering amplifier operates in two phases: reset phase and amplification phase. In the reset phase (Figure \ref{figure1}-(a)), two capacitors at the output terminals ($C_D$) are precharged to $V_{DD}$ using two precharge ($PREC$) switches, while the tail capacitor $C_T$ is reset to zero by turning on the $RESET$ switch. At the same time, the two transistors $M_1$ and $M_2$ are turned off in order to block charge flow between the capacitors. In the amplification phase (Figure \ref{figure1}-(b)), the $PREC$ and $RESET$ switches are turned off, and both transistors $M_1$ and $M_2$ are turned on with a differential input signal $V_{in} = V_{in(+)}-V_{in(-)}$. The tail capacitor $C_T$ is connected to the differential pair $M_1$ and $M_2$ through a switch enabled by the clock signal $CK$. During the amplification phase, charges in output capacitors are discharged to the tail capacitor for a certain period, and the relative amount of discharge from two output capacitors are controlled by the differential input signal. Therefore, a differential voltage gain $A_V$ is mainly determined by the ratio between the capacitance of $C_T$ and $C_D$ \cite{razavi2013charge}, with little dependence on the common mode voltage of $V_{in}$ as  
\begin{equation}
	A_v \approx \frac{2C_T}{C_D}.
	\label{equation1}
\end{equation}
Hence, a differential output signal $V_{out}$ is written as 
\begin{equation}
	V_{out}	\approx V_{in} \times A_v. 
	\label{equation2}
\end{equation}
where the $V_{out}$ is the differential output voltage between the $V_Q$ and $V_{QN}$. 
Consequently, a multi-level $V_{out}$ is generated as a product of two input variables: differential input voltage $V_{in}$ and amplifier gain $A_v$ (or $C_T$).

The charge-steering amplifier has two advantages over a traditional current-steering amplifier. First, its discrete operation is compatible with other digital circuits and helps save unnecessary power consumption. Secondly, it maintains stable operation even at high operating frequency up to a few GHz domains\cite{jung201325,chiang201410}. These render the charge-steering topology also suitable for analog MAC operations where both the power efficiency and speed are required.
\subsection{Mapping a Charge-Steering Amplifier onto Two 1T1C Cells}
\begin{figure}[h]
	\centering
	\includegraphics[width=0.85\columnwidth]{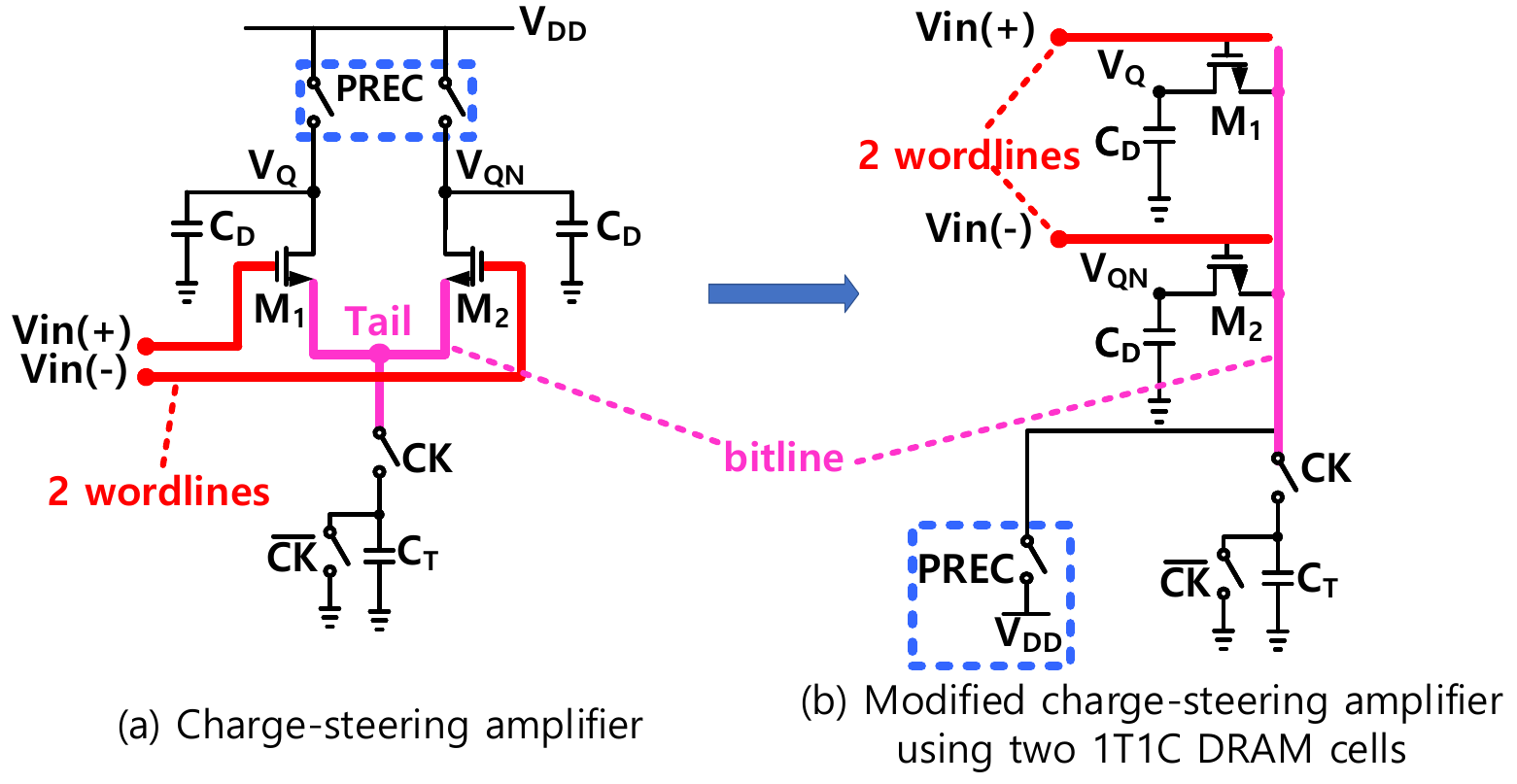}
	\caption{A modified charge-steering amplifier by using two 1T1C DRAM cells}
	\label{figure11}
\end{figure}
\begin{figure}[!t]
	\centering
	\includegraphics[width=0.9\columnwidth]{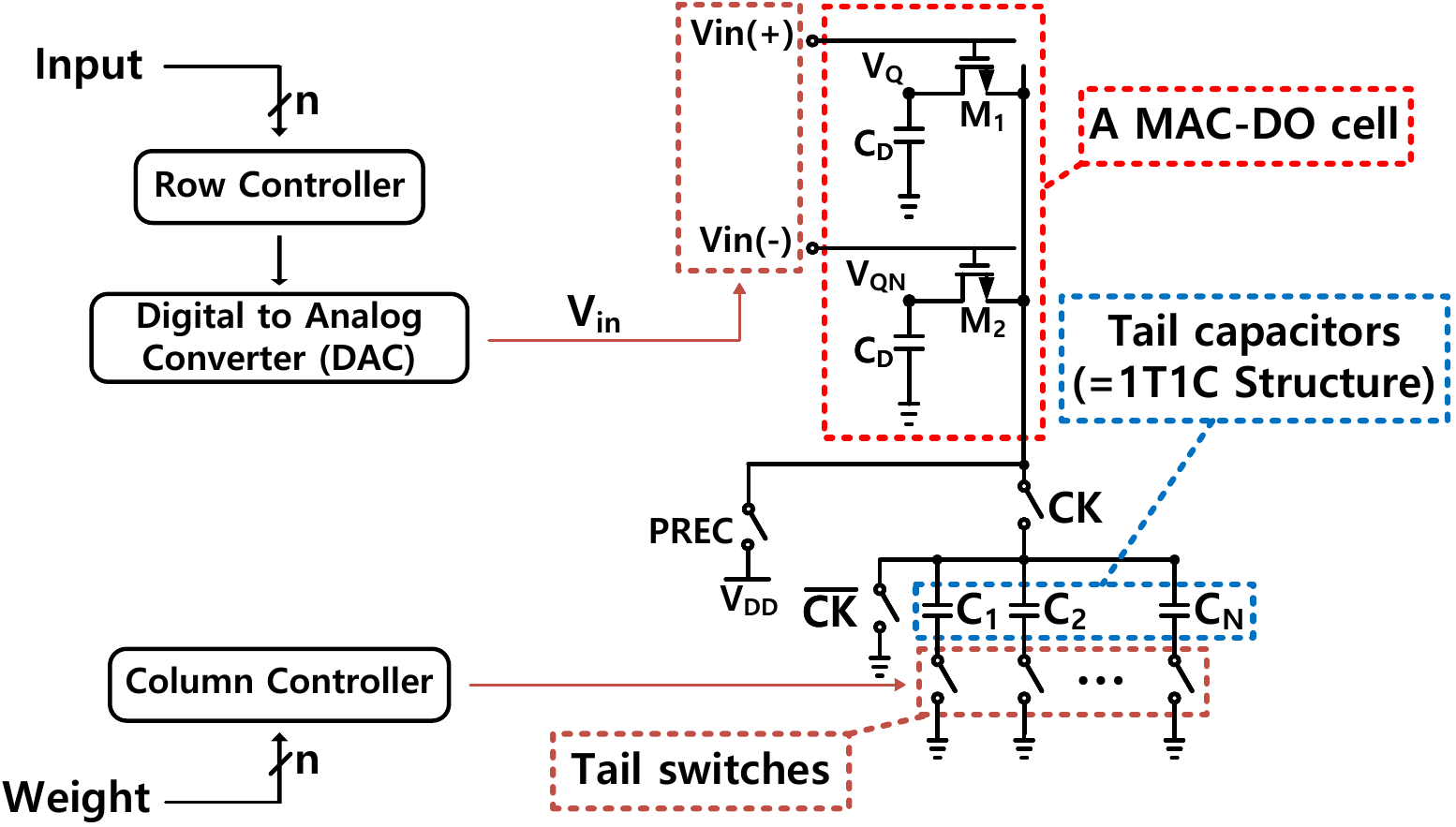}
	\caption{A MAC-DO cell for multi-bit MAC operations with peripheral circuits}
	\label{figure2}
\end{figure}
\begin{figure*}[t]
	\centering
	\includegraphics[width=0.85\linewidth]{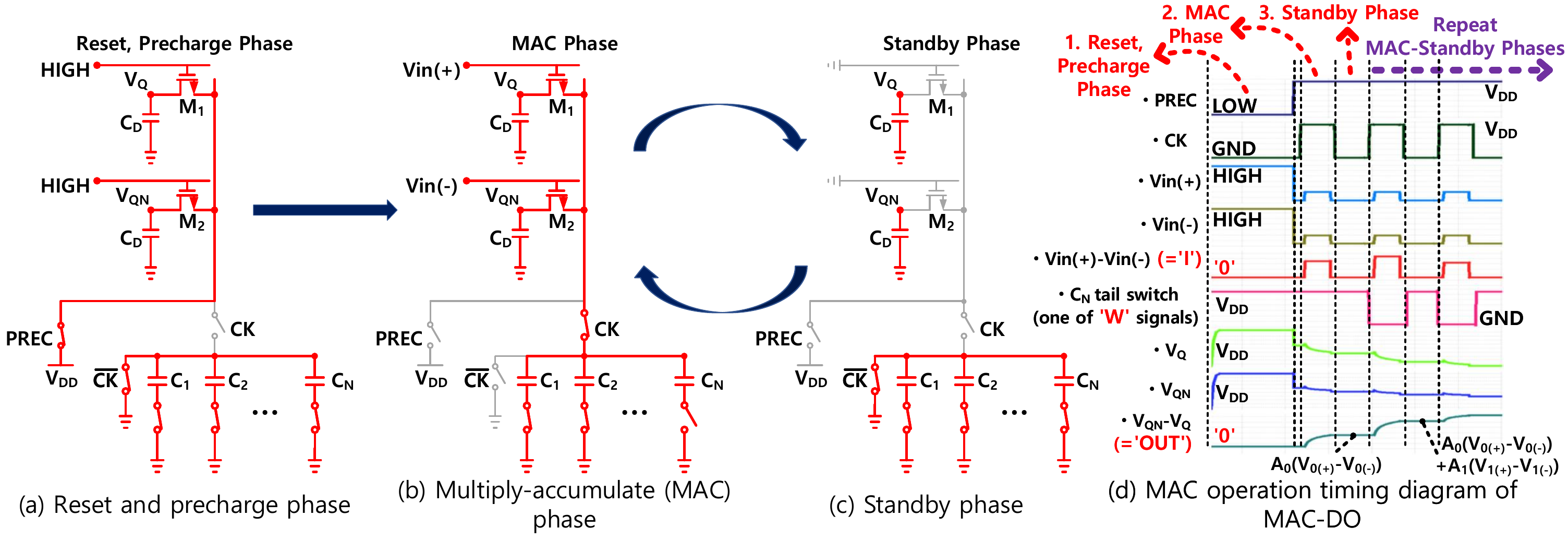}
	\caption{MAC operation phases of MAC-DO}
	\label{figure3}
 \vspace{-2em}
\end{figure*}

A DRAM array can be reorganized into an array of charge-steering amplifiers, or MAC-DO cells. Figure \ref{figure11}-(b) shows a modified charge-steering amplifier mapped on a DRAM array, which consists of two access transistors and two cell capacitors (two 1T1C DRAM cells). The tail node of the differential pair $M_1$ and $M_2$ in the charge-steering amplifier corresponds to the bit-line of the two 1T1C DRAM cells. Two $PREC$ switches in the original charge-steering amplifier are combined with $M_1$ and $M_2$ with an extra $PREC$ switch at the tail node, or the corresponding bit-line as shown in Figure \ref{figure11}-(b). In this modified charge-steering amplifier, $V_Q$ and $V_{QN}$ are precharged to $V_{DD}$ by turning on $M_1$, $M_2$, and $PREC$. $M_1$ and $M_2$ transistors need to be turned on by a voltage higher than $V_{DD}+V_{TH}$ in order to fully precharge two DRAM cell capacitors to $V_{DD}$. During amplification, two WLs of the modified charge-steering amplifier receive $V_{in(+)}$ and $V_{in(-)}$ voltages composing a differential input $V_{in}$. 
Since Figure \ref{figure11}-(a) and Figure \ref{figure11}-(b) are identical except the position and the number of $PREC$ switches, the modified charge-steering amplifier follows the same operation phases as discussed in Section \uppercase\expandafter{\romannumeral3}-A. As a result, a differential output signal $V_{out}$ is generated from the two DRAM cells, at $V_Q$ and $V_{QN}$.

\subsection{A MAC-DO Cell for a Series of Multi-Bit MAC Operations}
In order for the modified charge-steering amplifier to perform MAC operations with multi-bit input and weight data, $V_{in}$s and $A_v$s must be controllable by the input and weight data. It requires modifications on wordline and bitline drivers to the DRAM array MAC-DO cells are mapped on. In addition, a series of accumulations for an output stationary data flow involves a small change of operation phases.

\subsubsection{A MAC-DO cell and WL/BL drivers} A MAC-DO cell consists of two 1T1C DRAM cells as shown in Figure \ref{figure2} and carries out multi-bit MAC operations within the cell. A multi-bit digital input is converted into a differential input signal $V_{in}$ for $M_1$ and $M_2$ through a digital to analog converter (DAC). A multi-bit digital weight controls effective capacitance of the tail capacitor $C_T$ by enabling a part of parallel tail switches through a thermometer code decoder, and therefore controls the gain of the amplifier, $A_v$. Here, a MAC-DO cell is located inside a DRAM array and the other circuits are in the array periphery. In addition, a tail switch and a tail capacitor, which can be manufactured using DRAM technology, also follow the 1T1C structure.
Each multiplication result keeps accumulated at the $V_Q$ and $V_{QN}$ as a differential signal $V_{out}$ without additional precharge phases. This innate accumulation process necessitates an output stationary data flow for the array control since the MAC-DO cell is optimized for accumulating MAC outputs instead of storing input and weight data. The detailed MAC operation consists of three phases as follows.

\subsubsection{Phase 1. Reset and precharge phase} In the first phase, a MAC-DO cell is prepared for following MAC operations as shown in Figure \ref{figure3}-(a). The $PREC$ switch, $M_1$ and $M_2$ are turned on, so that two cell capacitors are fully precharged to $V_{DD}$, resulting in $V_{out}$=$V_{QN}$-$V_Q$=0. At the same time, the $RESET$ switch and all tail switches are turned on to reset all tail capacitors. These two operations are independent and both are carried out in this phase.

\subsubsection{Phase 2. Multiply-accumulate (MAC) phase} In this phase, the $PREC$ and $RESET$ switches are turned off first. Then, a differential $V_{in}$ signal corresponding to a multi-bit input is applied to $M_1$ and $M_2$ through a DAC. The $A_v$ value is adjusted according to a multi-bit weight by controlling the tail switches through a thermometer code decoder. As a result, the MAC-DO cell performs multiplication of multi-bit input and weight and generates a differential output voltage $V_{out}$ as
\begin{equation}
	V_{out} = V_{in}\times\sum_{i=1}^{N}\frac{2C_i}{C_D}
	\label{equation3}
\end{equation}
, and the multiplication result $V_{out}$ is accumulated at two cell capacitors as a differential voltage. Here, N determines the ratio of $\frac{C_T}{C_D}$ and hence the differential gain $A_v$. For example, if $N$ is 2, two tail switches are turned on as shown in in Figure \ref{figure3}-(b) and it leads to $V_{out}$ = $V_{in}$$\times$$\frac{2(C_1+C_2)}{C_D}$. Higher $N$ increases $A_v$ and is used for greater weights. With a proper conversion between the analog and digital domains, the Equation (\ref{equation3}) can be transformed as 
\begin{equation}
    \vspace{-1em}
	OUT = I\times W 
	\label{equation4}
\end{equation}
where $OUT$, $I$ and $W$ corresponds to $V_{out}$, $V_{in}$ and $\sum_{i=1}^{N}\frac{2C_i}{C_D}=A_v$, respectively. Both the input (I) and weight (W) can be easily converted to corresponding analog values ($V_{in}$ and $A_v$) by using a DAC for $V_{in}$ and a bank of tail capacitors for $A_v=\sum_{i=1}^{N}\frac{2C_i}{C_D}$. 

\subsubsection{Phase 3. Standby phase} After a MAC operation is performed in the MAC phase, $M_1$ and $M_2$ are turned off. Accordingly, the MAC result $V_{out}$ is stored at two cell capacitors as a differential analog voltage (Figure \ref{figure3}-(c)). Meanwhile, all tail switches and the $RESET$ switch are turned on to reset all tail capacitors for the next MAC operation.

A final MAC result is obtained by repeating only the MAC phase and the standby phase alternately without additional precharge phases, retaining the previous MAC result at $V_Q$ and $V_{QN}$. For instance, when next input and weight data are applied after the standby phase, the new multiplication is performed (Figure \ref{figure3}-(b)) and the result is accumulated in the same capacitors (at $V_Q$ and $V_{QN}$) with the previous MAC result (output stationary). Output voltages stored in the cell capacitors barely affect the new multiplication because they are connected to the drains of the differential pair. Therefore, the final MAC result in the MAC-DO cell after a series of the MAC-standby phases can be expressed as
\begin{equation}
	OUT_{FINAL}=\sum_{i}OUT_{i} = \sum_{i}I_{i}\times {W_i} 
	\label{equation5}
\end{equation}
which is the same equation as a vector dot product operation, or a series of multiply-accumulate (MAC) operations.

Figure \ref{figure3}-(d) shows the detailed timing diagram for MAC operations of a MAC-DO cell. Since the MAC-DO cell performs a series of MAC operations without additional precharge phases once cell capacitors are precharged to $V_{DD}$ in the beginning, it features an outstanding energy efficiency.

\begin{figure}[!t]
	\centering
	\includegraphics[width=\columnwidth]{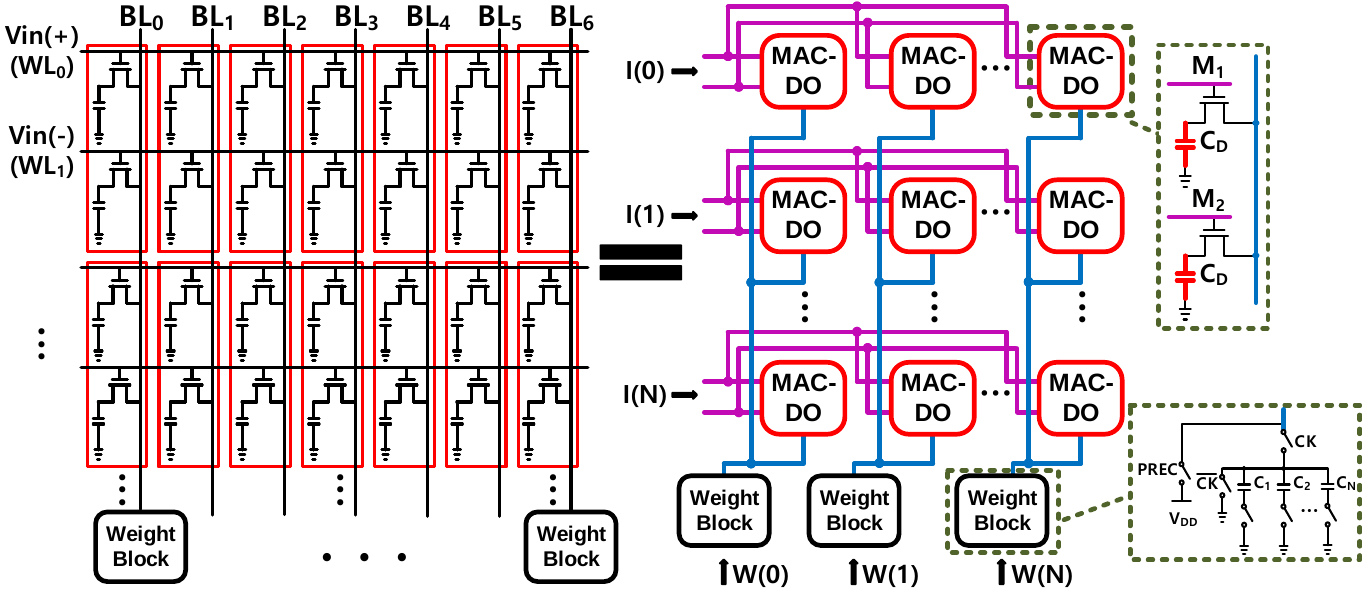}
	\caption{A MAC-DO array structure}
	\label{figure12}
 \vspace{-1em}
\end{figure}

\begin{figure}[!t]
	\centering
	\includegraphics[width=0.7\columnwidth]{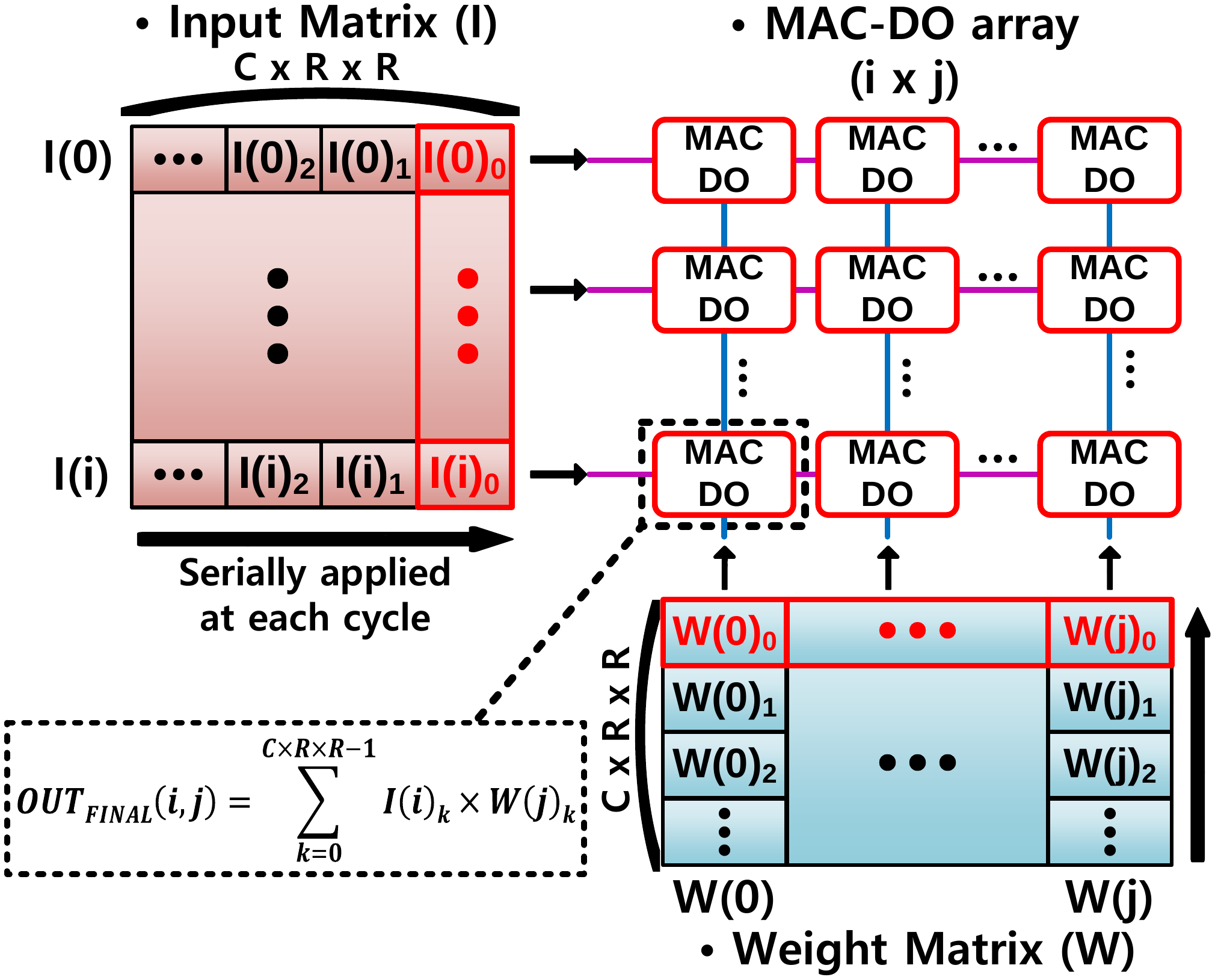}
	\caption{Matrix multiplications on a MAC-DO array for convolutions}
	\label{figure31}
\end{figure}

\subsection{Array Structure}

Multiple MAC-DO cells are combined and reorganized into an array for computing highly parallel MAC operations. A MAC-DO array is basically same as a DRAM array as shown in Figure \ref{figure12}. An input activation (I) is converted into a differential input voltage $V_{in}$ and shared across a row of MAC-DO cells using two WLs. Similarly, a weight (W) controls a tail capacitor bank (=weight block) added to a bit-line and $A_v$, which is shared across a column through the bit-line. Now, $A_v$ of each MAC-DO cell is $A_v \approx \frac{2C_T}{N\times C_D}$, where N is the number of MAC-DO cells in a column. Thanks to these input and weight broadcasting and output stationary, the MAC-DO array can calculate the outer product of two vectors ($I$ and $W$) and accumulate the result in the array at every cycle. In this way, a MAC-DO array can efficiently calculate the product of two matrices. For input and weight matrices with proper sizes, every MAC-DO cell inside the array can be engaged in individual MAC operations without leaving an idle cell. Therefore, the MAC-DO array architecture has a high utilization ratio of up to $100\%$ and high throughput compared to previous DRAM-based in-situ accelerators \cite{deng2018dracc,li2017drisa,seshadri2017ambit,angizi2019redram,shin2019pvt,li2018scope,roy2021pim}.

\subsection{Matrix Multiplications using a MAC-DO array}
Since each MAC result keeps accumulated in each MAC-DO cell, the MAC-DO array is controlled for an "output stationary" data flow \cite{moons201714,du2015shidiannao,reagen2021cheetah}. To process a CNN layer using the MAC-DO array, input matrix ($I$) and weight matrix ($W$) are prepared as shown in Figure \ref{figure31}. Then, the two matrices are multiplied through iterative outer products on the MAC-DO array. Each outer product is mapped on the MAC-DO array and keeps accumulated in each MAC-DO cell, resulting in an output activation
\begin{equation}
	\sum{OUT}(i,j) = \sum_{k=0}^{C \times R\times R-1}I(i)_k\times W(j)_k
	\label{equation6}
\end{equation}
where $i$ and $j$ represent the row and column number of the MAC-DO array, respectively, and k denotes the N'th computation cycle. After the matrix multiplication is completed, every MAC-DO cell stores an individual MAC result simultaneously. This capability enables MAC-DO to perform various convolution operations efficiently.

\subsection{Data Movement of a MAC-DO array}
The "output stationary" array control provides several benefits in terms of data movement. Once input and weight data are fetched from other memory arrays, they are shared across each row and column of the MAC-DO array and reused as many time as the array size. Also, the MAC results are stationary within each MAC-DO cell for the entire matrix multiplication cycles. Besides, a single cycle MAC operation within a MAC-DO cell minimizes the overall data access. Thanks to these efficient data reuse for all three types of data and a single cycle MAC operation, the MAC-DO array can efficiently minimize the data movement cost compared to previous DRAM-based in-situ accelerators \cite{deng2018dracc,li2017drisa,seshadri2017ambit,angizi2019redram,shin2019pvt,li2018scope,roy2021pim}.

\begin{figure}[t]
	\centering
	\includegraphics[width=0.7\columnwidth]{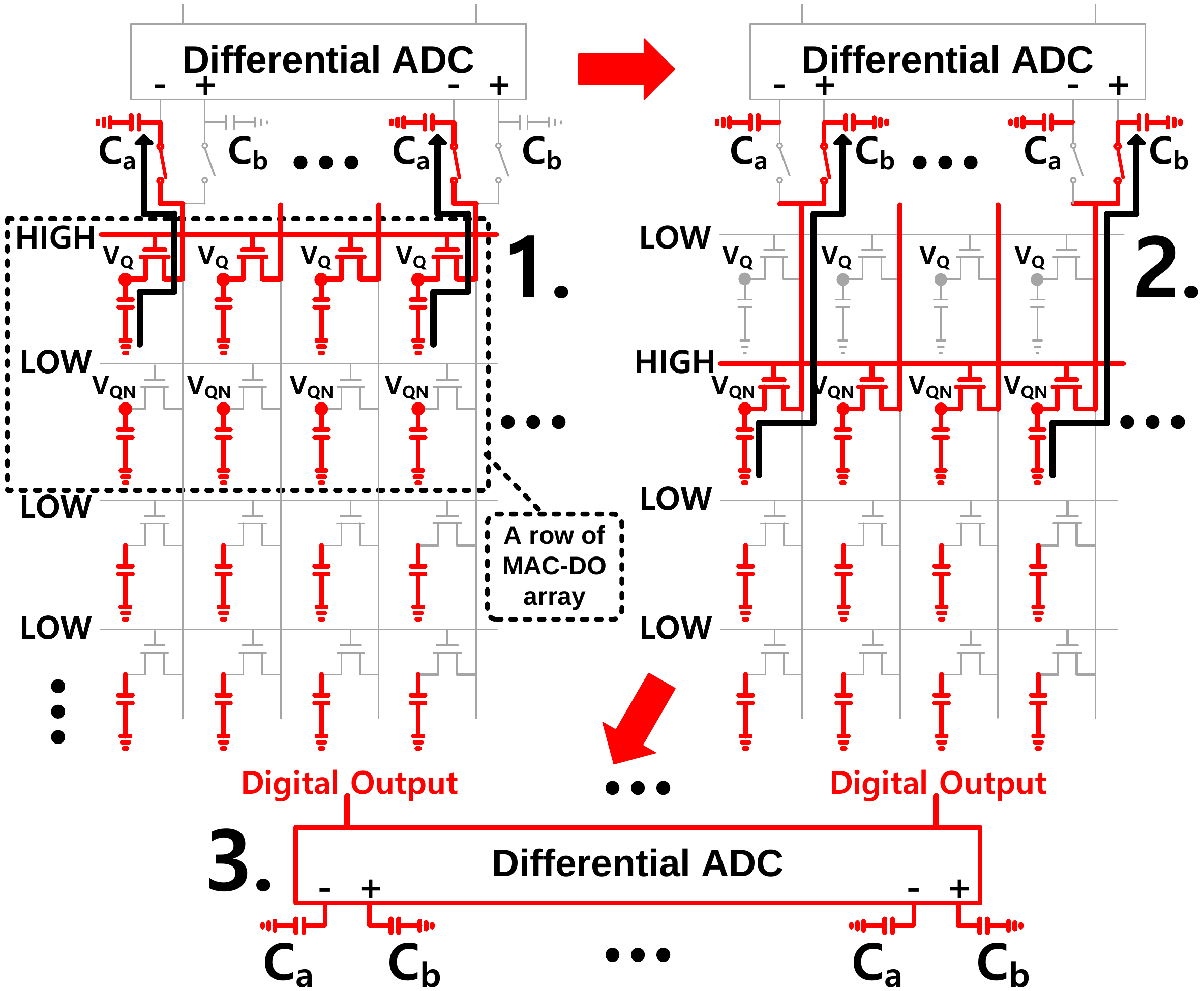}
	\caption{ADC conversion of MAC results}
	\label{figure28}
\end{figure}

\subsection{Reading out MAC Results}

Since each MAC-DO cell in the MAC-DO array stores an individual MAC result as a differential analog voltage across its two cell capacitors, a dedicated analog-to-digital converter (ADC) is required for reading out the stored results. The ADCs are connected to the MAC-DO array bitlines and quantize the analog MAC results. During the readout process, only one row of the MAC-DO array is involved in ADC conversion at a time, while the other rows are deactivated.
The readout process proceeds as follows: First, one of the two wordlines (WLs) of the row is activated, and the analog voltages stored at the $V_Q$ nodes are sampled on a row of $C_a$ capacitors and then held (Sample-and-Hold, S/H). Next, the other WL is activated to sample the other part of the differential voltages at $V_{QN}$ on a row of $C_b$ capacitors for the differential pair. Each ADC then quantizes the differential analog voltage sampled on two capacitors ($C_a$ and $C_b$). Once the MAC results for one row are read out, the process continues for the next rows in a row-wise fashion. Figure \ref{figure28} illustrates this readout process.

\subsection{Supporting Signed Number Operations}

\begin{figure}[h]
\vspace{-1em}
	\centering
	\includegraphics[width=0.5\columnwidth]{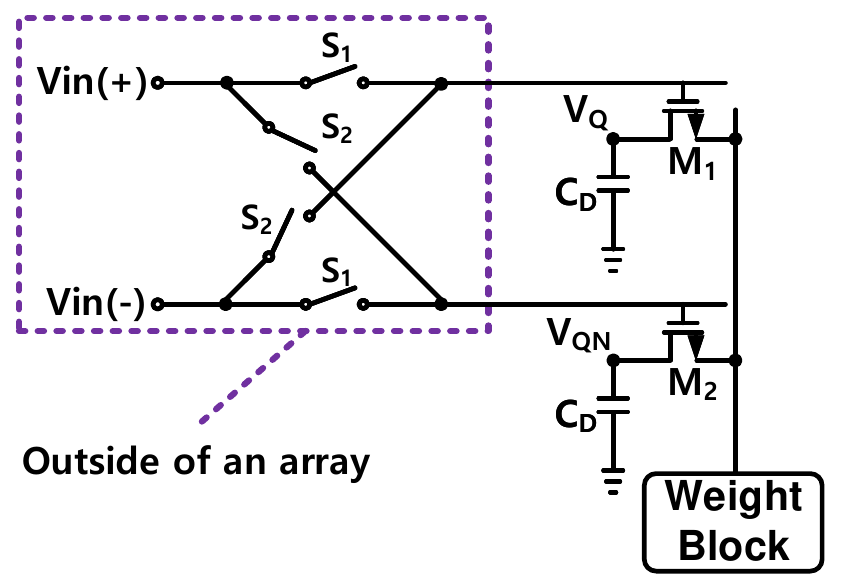}
	\caption{Signed bit operations for input activation}
	\label{figure7}
\end{figure}

\subsubsection{Signed input} Because an input activation is translated into a differential analog voltage, MAC-DO easily supports negative activation by flipping its polarity. This requires only a few additional switches as shown in Figure \ref{figure7}. The circuit is unchanged when $S_1$ switches are on. On the other hand, when $S_2$s are on, the polarity of the differential input signal $V_{in}$ is inverted inside the corresponding MAC-DO cells and a multiplication with negative input
\begin{equation}
	OUT = -I\times W 
	\label{equation8}
\end{equation}
occurs in the MAC-DO cells. This adds an extra sign bit for input and increases the bit precision.

\subsubsection{Signed weight} The charge-steering circuit always discharges from the DRAM cell capacitors, so a MAC-DO cell itself cannot handle negative or zero weight data. Also, because of the innate tail capacitance offset arising from parasitic capacitors at the BL and capacitor bank, the weight term (W) in Equation (\ref{equation4}) is biased and needs correction. In order to resolve both issues, a digital offset is added to weights before going into the array. So, the Equation (\ref{equation5}) for MAC operation is modified as
\begin{equation}  
\begin{split}
    \sum{OUT} = \sum{I\times (W+W_o+2^{N-1})} \\
    = \sum{I\times (W+W_c)}  
    \end{split}
\label{equation9} 
\end{equation}
where $W_o$ is the offset from parasitic capacitors, $N$ is the weight bit-precision including a signed bit and $2^{N-1}$ is a digitally added value for shifting negative weights into positives.

\section{Non-linear Effects Correction Methods}

\subsection{Mismatch Effect of MAC-DO Cells}

\begin{figure}[t]
	\centering
	\includegraphics[width=0.72\columnwidth]{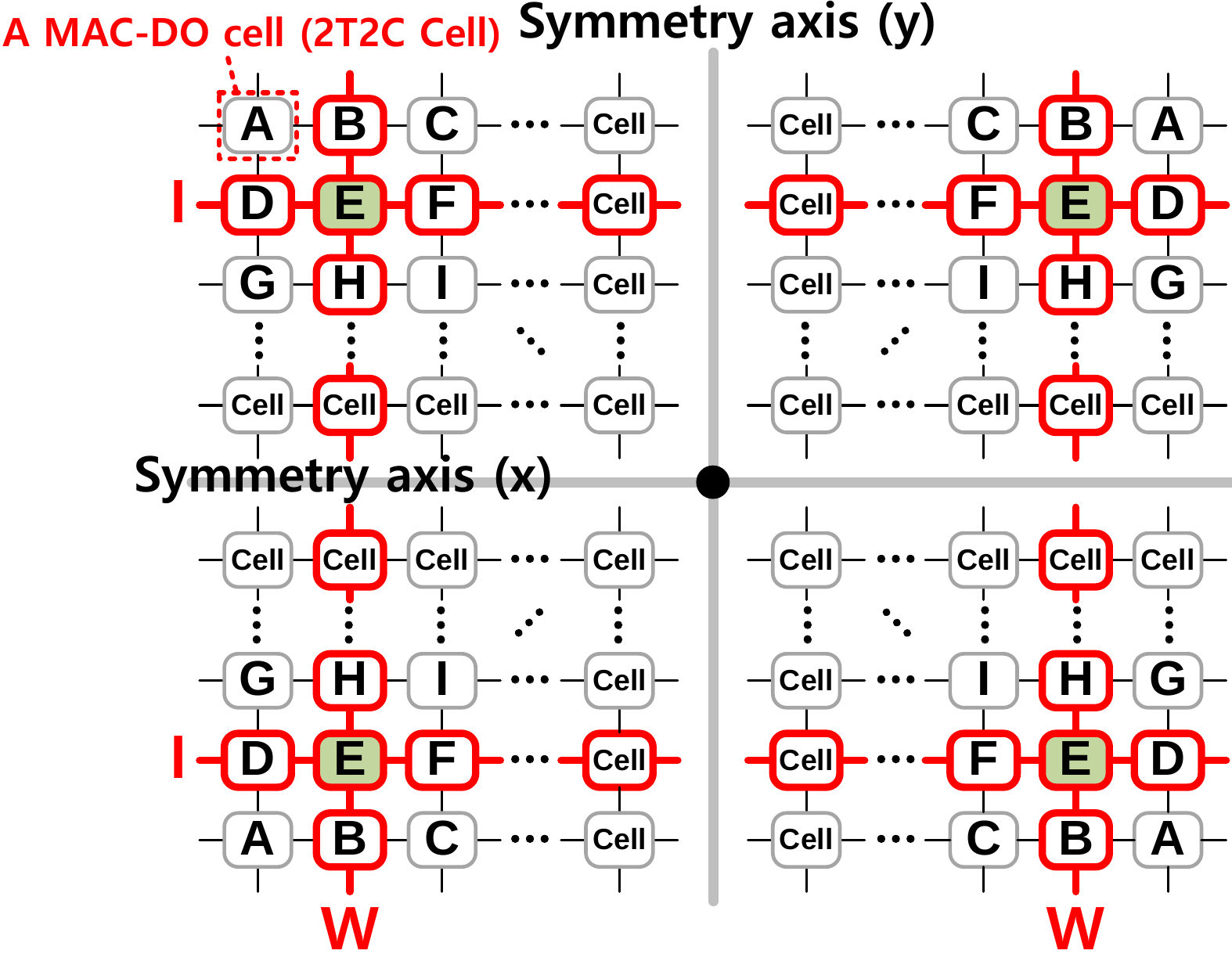}
	\caption{Common centroid layout for reducing mismatch effect}
	\label{figure18}
\end{figure}

Since every access transistor in a MAC-DO array performs analog MAC operations, the mismatch among the access transistors affects MAC operations in a real chip. As a result, outputs generated in two MAC-DO cells can be different even with the same input and weight data. To minimize the mismatch effects in MAC operations, three methods are employed in MAC-DO. Firstly, increasing the size of cell transistors reduces mismatch. Even though this increases the power consumption for driving cell transistors, it still maintains good energy efficiency. Secondly, a common centroid layout technique \cite{bastos1996matching} is used, symmetrically duplicating MAC-DO cells around both axes (x and y) within the DRAM array (Figure \ref{figure18}). These duplicated cells operate simultaneously with identical data, effectively reducing spatial mismatch gradients with minimal cell area increase. For example, 4 cells of 'E' in Figure \ref{figure18} are placed symmetrically inside a DRAM array and operate simultaneously. 

Despite those two solutions for reducing the mismatch effect, it is not fully removed in reality. With the mismatch effect, the Equation (\ref{equation9}) is expressed as
\begin{equation}
        \sum{OUT} = \sum{(I+I_m)\times (W+W_c)} 
    \label{equation11} 
\end{equation}
where $I_m$ represents the offset mismatch of each MAC-DO cell. 
Thus, MAC-DO requires additional correction techniques to cancel the offset terms ($I_m$ and $W_c$) to acquire an actual MAC result, which is expressed as $\sum{I\times W}$.

\subsection{Digital Correction}
To get the desired MAC result $\sum{I\times W}$, the left and right sides of Equation (\ref{equation11}) are transposed as
\begin{equation}
    \sum{I\times W}=\sum{OUT}-I_{m}\sum{W}-W_{c}\sum{I}-\sum{I_{m}W_c}
	\label{equation12} 
\end{equation}
, so the offset effects included in $\sum{OUT}$ need to be subtracted to get the desired MAC result. The offset constants of $I_m$ and $W_c$ are obtained by applying the test data composed of '1' and '0' and by solving the equation above. For this correction, MAC-DO needs additional accumulations of input and weight data in the digital domain, but its overhead is not critical in the entire system, since the offset constants can be reused once they are obtained and other accumulation results can also be shared across many cells in a row or column.

\subsection{Analog Correction} 
In addition to the digital correction, MAC-DO can use an analog offset cancellation technique such as chopping \cite{van20012}. For this, MAC-DO performs an additional MAC operation with negated input and weight after a normal MAC cycle. The two MAC operation results, $OUT$ and $OUT'$, are
\begin{equation}  
    \begin{split}
    OUT = (I+I_m)\times(W+W_c) \\
    OUT' = (-I+I_m)\times(-W+W_c),
    \end{split}
	\label{equation13}
\end{equation}
and they add up to
\begin{equation}  
    OUT+OUT' = 2(I \times W + I_m \times W_c).
	\label{equation14} 
\end{equation}
Therefore, the desired MAC result is expressed as
\begin{equation}  
    \sum{I\times W}=\left(\sum{OUT+OUT'}-\sum{I_{m}W_{c}}\right)/2
\label{equation15} 
\end{equation}
Now the MAC result has only one constant subtraction term that can be easily found and computed; It no longer requires accumulation of input nor weight data.

\subsection{Leakage Effects on MAC-DO Cells}
The MAC-DO cells may experience charge leakage as they store MAC results on DRAM cell capacitors during MAC operations.
However, the leakage effect is considered negligible due to the short duration of MAC-DO's operations (up to GHz, as described in Section \uppercase\expandafter{\romannumeral3}-A) within the DRAM refresh period (typically 64ms). Additionally, the leakage effects can be cancelled out in MAC-DO cells, as each cell reads the voltage difference between two adjacent output capacitors, where the leakage tends to have a similar tendency.
Furthermore, adopting high Vth transistors for DRAM access transistors or applying a negative voltage on word-lines during the standby phase can further mitigate leakage. Finally, during training, the leakage effect can be effectively addressed through appropriate modeling techniques \cite{joksas2022nonideality}.

\section{Evaluation Methodology}

\subsection{Overall System Architecture for Testing MAC-DO}
\begin{figure}[t]
	\centering
	\includegraphics[width=0.84\linewidth]{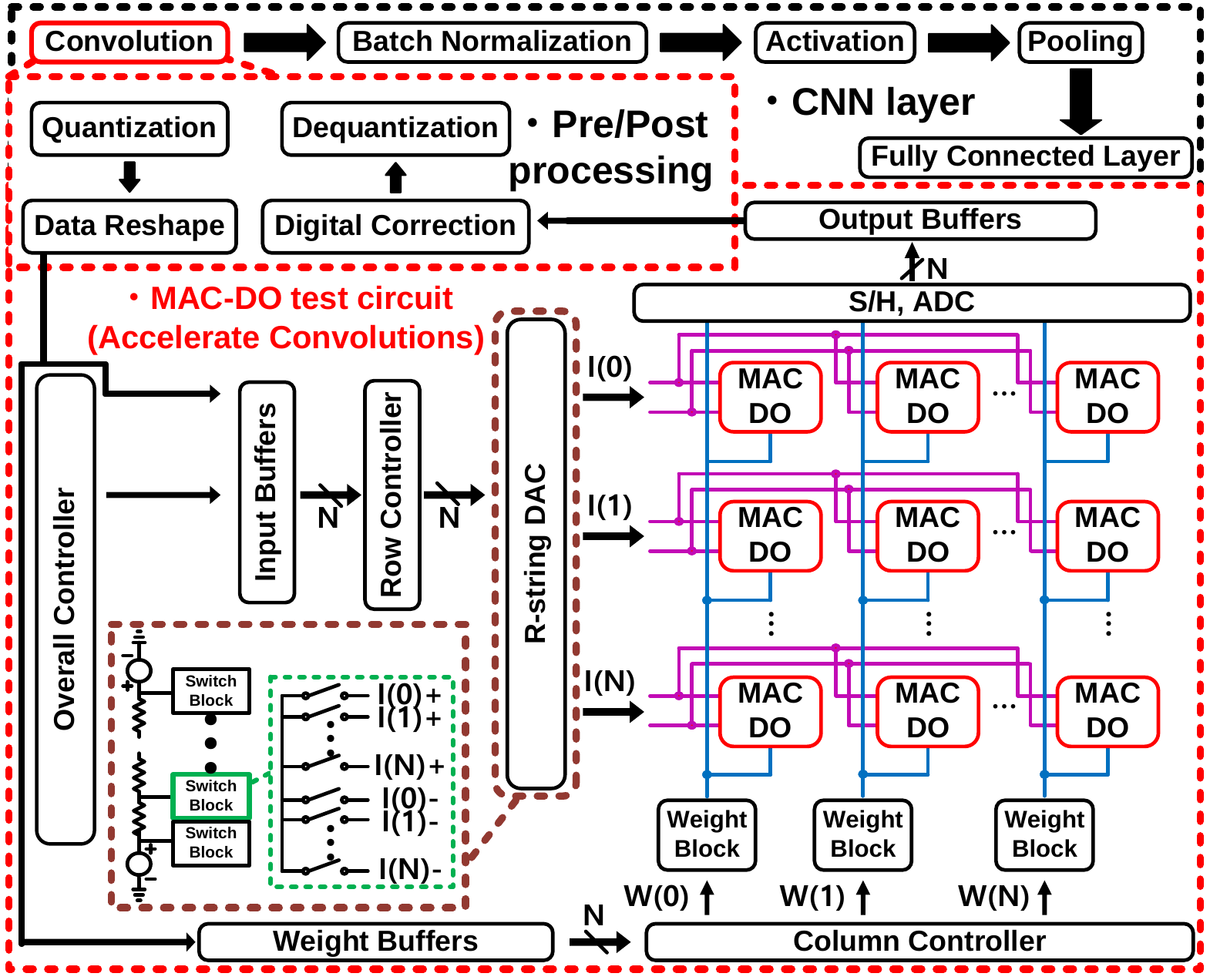}
	\caption{Overall system architecture for testing MAC-DO}
	\label{figure4}
\end{figure}

To verify the performance of MAC-DO, an overall system architecture including test circuits has been designed as shown in Figure \ref{figure4}. The test focuses on accelerating compute-intensive convolutions in inference which are the largest bottleneck in CNN layers ($>90$\% computations, runtime). The MAC-DO for accelerating convolutions mainly consists of five blocks: a MAC-DO array, a row controller (Row\_C), an R-string DAC block, a column controller (Col\_C), and an ADC block. 

For convolutions, input and weight data are stored outside of MAC-DO, which would be other memory arrays. Those data are quantized and go through a data-reshaping step to be prepared as a matrix form to perform matrix multiplications. Then, the row controller receives the input matrix through input buffers and controls the R-string DAC to generate differential analog input voltages for corresponding rows of MAC-DO cells. The column controller receives the weight matrix through weight buffers and manipulates the tail switches of the capacitor banks in weight blocks. After a series of MAC operations inside the array, the ADCs convert the differential analog output voltages stored at each MAC-DO cell into digital values row by row. Then, digital correction is performed and the digital values are dequantized. Data pre-/post processing and running other layers such as pooling are performed with the help of pytorch software \cite{paszke2017automatic} because they are rather data-intensive and can be easily performed using conventional in-/near-DRAM processing techniques. Detailed software/system support for end-to-end simulation is out of the scope of this paper and planned as future work.

\begin{scriptsize}
\begin{table}[h!]
    \vspace{-0.8em}
  \centering
  \caption{Design Parameters of MAC-DO Test Circuit}
  \label{table1}
  \renewcommand{\tabcolsep}{0.4mm}
  \begin{tabular}{|c|c|}
    \hline
    Technology & 65nm CMOS logic process \\
    
    \hline
    Supply voltage ($V_{DD}$) & 1.2V\\
    \hline
    Clock frequency & 12.5 (MHz)\\
    \hline
    An access transistor size (W/L) & 800/560 (nm)\\
    \hline
    Cell capacitance & 100f (F) \\
    \hline
    Noise at cell capacitors & \SI{264.3}{\micro\volt}, $<$ 0.13\% error\\
    \hline
    Leakage at cell capacitors & $\sim$~4\si{\nano\volt/\nano s}, $<$ 0.012\% error\\
    \hline
    Each tail capacitance  & 6.8f$\sim$9.6f (F)\\
    \hline
    A MAC-DO cell size & 221.21 (\SI{}{\micro\meter}$^2$)\\
    \hline
    An array size & 16 $\times$ 16\\
    \hline
    Input/weight precision & 4bit/4bit integer \\
    \hline
    \# of maximum MAC operations & 200\\
    in a MAC-DO cell          & (10.6 fJ/MAC) \\ 
    \hline
  \end{tabular}
  \vspace{-1em}
\end{table}
\end{scriptsize}
\begin{scriptsize}
	\begin{table}[!h]
		\centering
		\caption{DRAM parameters}
		\label{table11}
		\renewcommand{\tabcolsep}{0.4mm}
		\begin{tabular}{|c|}
			\hline
			  DRAM configuration \cite{jesd} \\
			\hline
                DDR4, 8Gb $\times$ 8, 4 bank groups, 4banks, 64K rows/bank, 1KB row \\
                technology = 22nm, area = 53.6mm$^2$, cell area = 0.0036um$^2$ \\
                \hline
		\end{tabular}
  \vspace{-1em}
	\end{table}
\end{scriptsize}

\subsection{Parameters of MAC-DO Test Circuits}
Major design parameters of the test circuit are shown in Table~\ref{table1}. The circuit has been designed using 65nm CMOS logic process and is powered by 1.2V supply voltage. Though the circuit can operate much faster, the clock frequency is set relatively slower to verify the robustness of the MAC-DO cell in one of the worse cases, as unwanted effects of leakage currents increase with longer period. The size of access transistors is selected larger than a usual 65 nm design to reflect the methods for reducing mismatch effects ($>$10$\times4$, 10 larger transistor size than normal design and 4 from common centroid). The MAC-DO array size is set as 16$\times$16 for only the test, and hence cell capacitance is chosen relatively larger than normal DRAM cell capacitance to adjust $A_v$. The error of output noise and leakage are less than 0.13\% and 0.012\%, respectively, out of the final MAC result. Each tail capacitance is sized to perform MAC operations with minimum linearity errors. Table \ref{table11} shows the DRAM parameters \cite{jesd} used as a baseline for evaluating the performance of MAC-DO.
The test circuit is optimized for 4b$\times$4b precision of input and weight data, considering non-linear effects and the area of MAC-DO because low-bit quantized inference is a common practice in edge devices \cite{choi2019optimized} due to their constraints, such as limited area and battery capacity. Additionally, various techniques exist to compensate for accuracy losses arising from low-bit precision \cite{mishra2017apprentice, choukroun2019low}. The test circuit's scalability can be achieved during the design phase by adjusting the DAC precision or the number of tail capacitors, depending on target noise levels and the area available for DAC and weight blocks in the array periphery. With these circuit parameters, a MAC-DO cell can perform a series of up to 200 MAC operations ($>$250mV) without precharging the output capacitors again. This can be further increased depending on the clock frequency, tail capacitance and levels of wordline input voltages $V_{in}$. 

\subsection{Circuit Level Implementation}
MAC-DO has been simulated in transistor level by using Cadence Spectre Simulator \cite{cadence}. The MAC-DO array (16$\times$16 MAC-DO cells or 32$\times$16 1T1C DRAM cells), row controller (including switch blocks), R-string DAC and column controller have been designed in transistor level by using Cadence Virtuoso Schematic Editor \cite{cadence1}. For ADC analysis, we use data from a survey of recent ADC circuits \cite{ADCcite} and scaled an ADC data to 65nm process, 1.2V supply voltage and 6bit output precision. We assume 16 ADCs (=\# of columns) are used and each ADC area is 0.00116mm{$^2$} with 0.89 (pJ) per 6bit conversions. The test focuses on accelerating convolutions of a neural network. The 4bit quantized input and weight data at each convolution layer are extracted through PyTorch\cite{paszke2017automatic} and they are transferred into the Spectre simulator by using a Verilog-A block modeling a data bus \cite{ fitzpatrick1998analog }. After a simulation is finished, differential analog output voltages of the simulated array are directly observed from every MAC-DO cell simultaneously and delivered to PyTorch to dequantize the analog MAC results and run remaining layers.

\subsection{Dataset and Network for MAC-DO Circuit Verification}
\begin{scriptsize}
	\begin{table}[h!]
		\centering
		\caption{LeNet-5 Neural Network for Circuit Simulation}
		\renewcommand{\tabcolsep}{1mm}
		\label{table2}
		\begin{tabular}{|c|c|}
			\hline
			Dataset, Network   & MNIST\cite{mnistdataset}, LeNet-5\cite{lecun1998gradient} \\

			\hline
			Batch size & 32 \\
			\hline
			\hline
			Layers     &    Network Parameters  \\
			\hline
			Conv1(C1) & Input Feature : 1 x 32 x 32 \\ 
			BatchNorm, Tanh          & Weight Filter : 6 x 1 x 5 x 5               \\

			\hline
			Conv3(C3) & Input Feature : 6 x 14 x 14 \\ 
			BatchNorm, Tanh          & Weight Filter : 16 x 6 x 5 x 5               \\

			\hline
			Conv5(C5) & Input Feature : 16 x 5 x 5 \\ 
			BatchNorm, Tanh		  & Weight Filter : 120 x 16 x 5 x 5               \\	
			
			\hline
			FC1       & Input Feature : 120 x 1 \\
				Tanh	  & Weight Filter : 84 x 120 \\
			
			\hline
			FC2       & Input Feature : 84 x 1 \\
					  & Weight Filter : 10 x 84 \\
			\hline			
		\end{tabular}
	\end{table}
\end{scriptsize}

To verify matrix multiplications on the MAC-DO array, MAC-DO's computation accuracy has been tested in the same way as \cite{jung2022crossbar}. We measure the Top-1 accuracy drops when MAC-DO accelerates a convolution layer of LeNet-5\cite{lecun1998gradient} neural network for MNIST dataset\cite{mnistdataset}. Other layers such as non-linear function have been supported by using software. The detailed network parameters of LeNet-5 is shown in Table~\ref{table2}.
For benchmarking, the LeNet-5 network for MNIST dataset is pre-trained with full precision operations and batch size 32 using PyTorch, and the Top-1 accuracy shows 99.075\%. Also, a convolution layer in LeNet-5 has been digitally computed after 4bit, 3bit and 2bit quantization to compare with analog computation using MAC-DO. For these tests, the pre-trained network has been reused without retraining and a digital correction is performed using PyTorch. The accuracy for the each case is shown in Table~\ref{table3}.

\begin{scriptsize}
	\begin{table}[h!]
		\centering
        \vspace{-0.8em}
		\caption{Benchmarking Top-1 Accuracy for MNIST Dataset in LeNet-5}
		\label{table3}
		\renewcommand{\tabcolsep}{0.5mm}
		\begin{tabular}{|c|c|c|c|c|}
			\hline
			I/W precision  & full-precision & 4b/4b & 3b/3b & 2b/2b \\
			\hline
			\hline
            Top-1 Accuracy & 99.075\% & 98.973\% & 98.595\% &84.767\% \\
			\hline			
		\end{tabular}
	\end{table}
 \vspace{-1em}
\end{scriptsize}

\subsection{System Level Evaluation}
We developed an in-house cycle-based simulator to evaluate the data movement and speedup achieved by MAC-DO compared to other DRAM-based in-situ accelerators. The simulator incorporates data extracted from the circuit simulation to ensure accuracy.
For performance verification, we tested various convolutional neural networks, including LeNet-5 \cite{lecun1998gradient}, MobileNet V1 \cite{howard2017mobilenets}, MobileNet V2 \cite{sandler2018mobilenetv2}, ShuffleNet \cite{zhang2018shufflenet}, and ResNet-18 \cite{he2016deep}. The simulator assesses system performance by accelerating all convolution layers in the CNNs using a mat and takes into account operation cycles and data movement, including data copy.
In the comparison, each accelerator's mat size is set to 512$\times$512 1T1C DRAM cells, and the bit precision is fixed at 4bit for both input and weight data.

\section{Evaluation Results}
\subsection{Accuracy of MAC Results of MAC-DO Cells}
\begin{figure}[h]
    \vspace{-0.8em}
	\centering
	\includegraphics[width=\columnwidth]{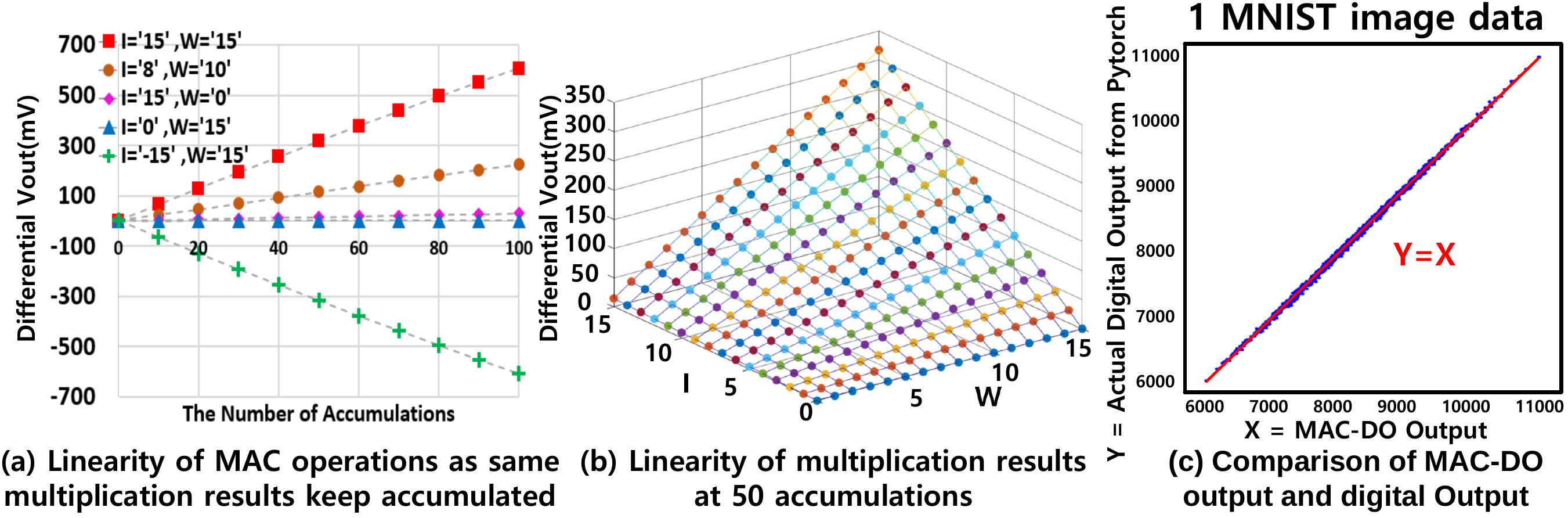}
	\caption{The linearity of multiplication results of a MAC-DO cell}
\label{figure13}
\end{figure}

Figure \ref{figure13} shows the accuracy of multiplication results of a MAC-DO cell. Figure \ref{figure13}-(a) shows the accuracy of repetitive MAC operations in a MAC-DO cell. For example, the plot with square markers shows the accumulation results ($V_{out}$) of a series of  15($I$)$\times$15($W$) multiplications. Results with W = 0, show that non-zero multiplication results keep accumulating because of the weight offset arising from parasitic capacitors in weight blocks. However, this unwanted offset can be removed by aforementioned correction techniques. Figure \ref{figure13}-(b) shows the multiplication results for all 256 (4b$\times$4b) combinations of inputs and weights, after 50 times of accumulation in a MAC-DO cell. Table \ref{table4} shows the relative error of the data in Figure \ref{figure13}-(b) from the ideal values with non-linear effects. Analog correction requires doubled MAC cycles but shows much better correction performance. Figure \ref{figure13}-(c) shows a comparison between MAC-DO's output and the actual digital output from PyTorch for one MNIST image.

\begin{scriptsize}
	\begin{table}[!h]
		\centering
    \vspace{-0.8em}
		\caption{Effects of Digital and Analog Mismatch Correction Methods}
		\label{table4}
		\renewcommand{\tabcolsep}{1mm}
		\begin{tabular}{|c|c|c|c|}
			\hline
			Correction  & No correction & Digital & Digital+Analog \\
			\hline
			\hline
            Error range(\%) & $\sim$4.06\% & $\sim$2\% & $\sim$0.23\%  \\
			\hline			
		\end{tabular}
  \vspace{-1em}
	\end{table}
\end{scriptsize}
\subsection{Inference Accuracy of MAC-DO}
The inference accuracy of MAC-DO has been tested with C3 layer of LeNet-5. For the test, C3 convolution layer is executed by the MAC-DO test circuit using transistor-level simulation for 448 test set images from the MNIST dataset, and the Top-1 accuracy is calculated from the collection of the final results. Other layers have been executed with full precision using software in the similar way as \cite{jung2022crossbar}. In order to dequantize the analog MAC results, we use four images as training data to find proper dequantization parameters. The Top-1 accuracy shows 97.07\% with a standard deviation of 0.2507\%, (without network retraining, the four training images are not included). To estimate an effective bit precision of this analog computing, the Top-1 accuracy is compared to the accuracy results when the same C3 layer only has been executed digitally after quantization into 2-, 3-, 4-bit data (Table \ref{table3}). The Top-1 accuracy drop of 1.903\% from the MAC-DO analog computing is most similar to that of digital operation with 3-bit quantized data. The accuracy number can be further improved by retraining the LeNet-5 network with MAC-DO circuits or performing additional analog corrections.

\subsection{Circuit Level Performance Analysis and Comparison}
\begin{figure}[h]
	\centering
 \vspace{-0.5em}
	\includegraphics[width=\columnwidth]{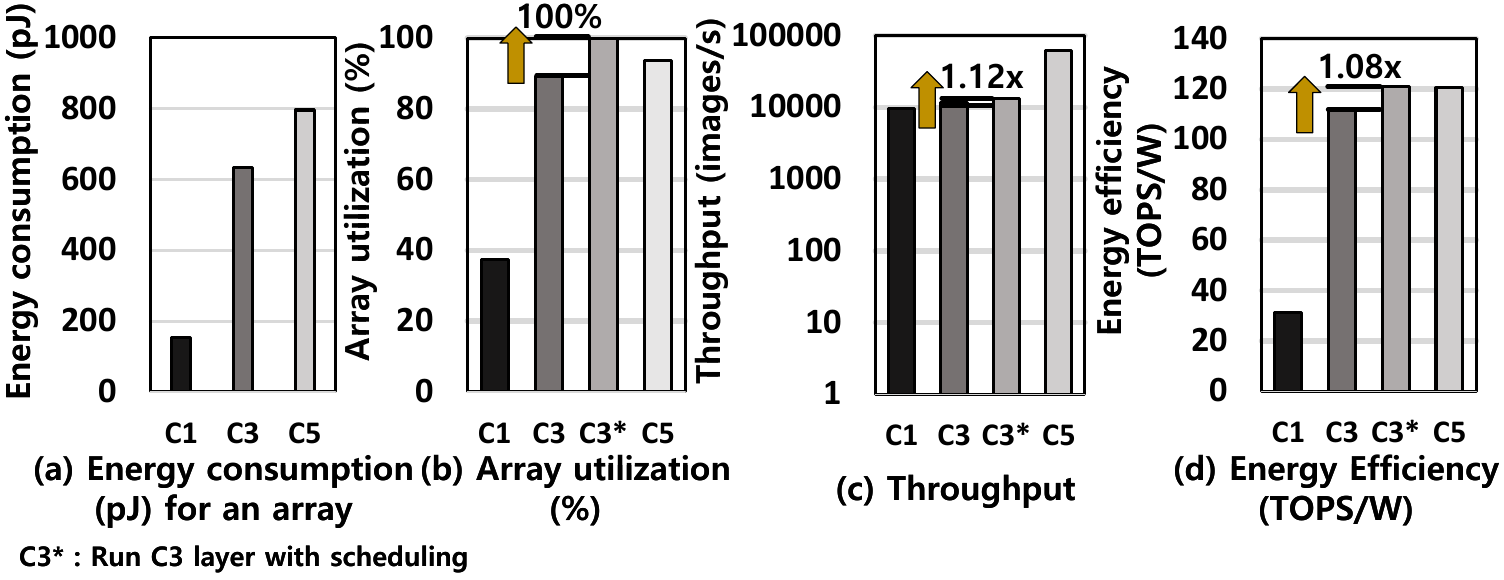}
	\caption{Performance comparison among convolution layers}
	\label{figure15}
\end{figure}

\begin{figure}[h]
	\centering
 \vspace{-0.5em}
	\includegraphics[width=0.95\columnwidth]{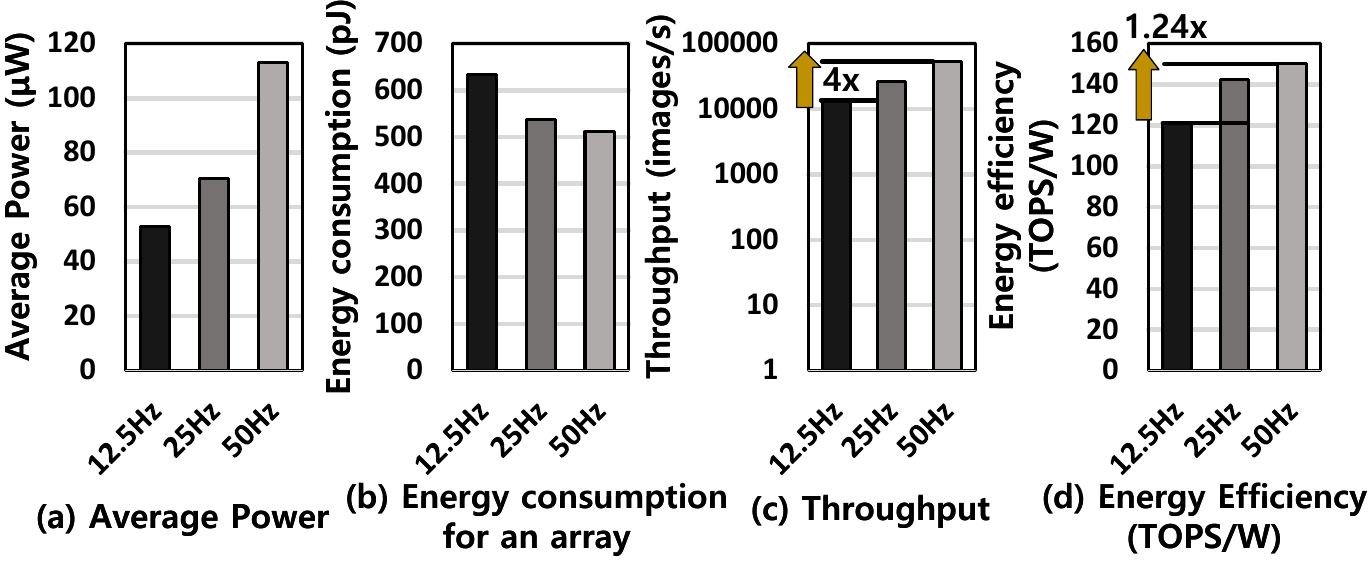}
	\caption{Performance comparison at faster speeds}
	\label{figure25}
\end{figure}

Figure \ref{figure15} summarizes the performance results from the circuit simulation of the MAC-DO test circuit on three convolution layers (C1, C3, C5) of LeNet-5. Figure \ref{figure15}-(a) displays the total energy consumption for executing an array operation (a portion of convolution that fits in the 16$\times$16 array). Figure \ref{figure15}-(b) indicates the average array utilization (used MAC-DO cells / total MAC-DO cells) for each convolution layer, with high utilization (approximately 93.75\%) except for the C1 layer, which has unusually few input and output channels (1 and 6, respectively). Figure \ref{figure15}-(c) represents the throughput in terms of images per second. Figure \ref{figure15}-(d) presents the energy efficiency (TOPS/W), with 1 MAC operation considered as 2 operations (1 multiply + 1 accumulate). C1 has the lowest array utilization, resulting in the lowest energy efficiency among the three layers.  
By scheduling and processing a part of image data with the next image, the array utilization can reach 100\% for the C3 layer (C3*), because the number of its output channels (=16) is a multiple of the number of columns (=16) in the MAC-DO array. 
In this case, the throughput and energy efficiency show 1.12$\times$ and 1.08$\times$ improvement over when without scheduling, respectively.
Figure \ref{figure25} illustrates the performance summary for faster clock frequency settings. The throughput increases linearly with the clock frequency because the time for precharging cell capacitors does not significantly affect precharging cycles due to the small array size. Additionally, it demonstrates better energy efficiency at faster speeds as excessive energy dissipation during too long evaluation periods is reduced, even though the average power increases.

\subsection{Average Power Breakdown of the MAC-DO Test Circuit}
\begin{figure}[!h]
	\centering
 \vspace{-0.5em}
	\includegraphics[width=0.8\columnwidth]{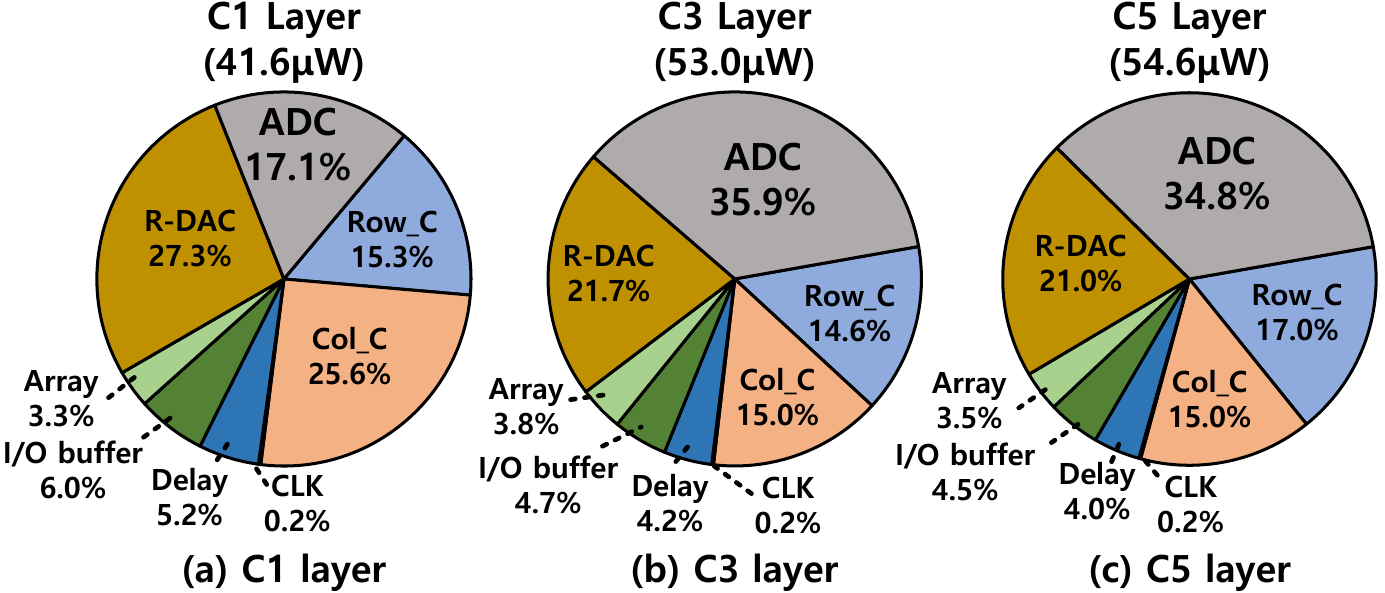}
	\caption{Average power breakdown of three convolution layers (C1, C3 and C5)}
	\label{figure14}
 \vspace{-1em}
\end{figure}

\begin{scriptsize}
	\begin{table}[!h]
		\centering
		\caption{Scaling of Average Power to 22nm DRAM Technology}
		\label{table9}
		\renewcommand{\tabcolsep}{0.4mm}
		\begin{tabular}{|c|c|c|}
			\hline
			Circuit & Power using CMOS tech  & Power using DRAM tech \\
			\hline
                R-DAC & \SI{11.43}{\micro\watt} & \SI{9.15}{\micro\watt} (0.8$\times$) \\
			\hline
                Row\_C & \SI{7.79}{\micro\watt} & \SI{3.14}{\micro\watt} (0.437$\times$) \\
			\hline
                Col\_C & \SI{8.92}{\micro\watt} & \SI{0.93}{\micro\watt} (0.104$\times$)\\
			\hline		
                Array & \SI{1.77}{\micro\watt} & \SI{0.17}{\micro\watt} (0.096$\times$)\\
                \hline
		\end{tabular}
	\end{table}
\end{scriptsize}

Figure \ref{figure14} shows the average power breakdown of the MAC-DO test circuit for running three convolution layers C1, C3 and C5 of LeNet-5. The total power for C1, C3 and C5 layers are \SI{41.6}{\micro\watt}, \SI{53.0}{\micro\watt} and \SI{54.6}{\micro\watt}, respectively. The power consumption for precharging cell capacitors is dominant in the array power because MAC operations in MAC-DO are performed by discharging cell capacitors that were precharged once at the precharge phase. The array power consumption in the C1 layer is smaller than in the C3 and C5 layers because it requires the fewest accumulation cycles for a convolution (5$\times$5 = 25 cycles). The ADCs account for large portion of the total power. The R-DAC also shows significant power consumption in the MAC-DO test circuit because it drives a lot of big access transistors ($\sim$40$\times$ bigger than usual), so it can be further reduced by optimizing the size of access transistors in DRAM process. In addition, since the size of a cell capacitor is smaller in actual DRAM process, the size of tail capacitors can also be scaled down, and so does the power consumption of the column controller (Col\_C) and the MAC array in a real DRAM based chip. Table \ref{table9} presents the scaling of average power for all convolution layers from CMOS technology to DRAM technology.

\subsection{Performance Estimation for a Real DRAM Based Array} 

\begin{scriptsize}
	\begin{table}[!h]
        \vspace{-1em}
		\centering
		\caption{Performance Estimation at Realistic Array Size}
		\label{table8}
		\renewcommand{\tabcolsep}{0.4mm}
		\begin{tabular}{|c|c|c|}
			\hline
			Power & Throughput  & Energy efficiency \\
			\hline
			\hline
             11.55 (mW) & 3.26 TOPS (509.4$\times$)  & 282.34 TOPS/W (2.33$\times$)\\
			\hline			
		\end{tabular}
	\end{table}
\end{scriptsize}
\begin{scriptsize}
	\begin{table*}[t!]
		\centering
		\caption{Baseline Descriptions for Other Works}
		\label{table6}
		\renewcommand{\tabcolsep}{1mm}
		\begin{tabular}{|c|c|c|c|c|c|c|c|c|c|}
			\hline
			&GPU & \multicolumn {2}{|c|}{Digital Accelerator}
			& \multicolumn {2}{|c|}{SRAM-Based CiM} & \multicolumn {4}{|c|}{DRAM-Based In-Situ Accelerator}
			\\
			
			\hline
			&TITAN-X\cite{TITAN}\footnote{1}& Eyeriss\cite{chen2016eyeriss}&DaDianNao\cite{chen2014dadiannao}\footnote{2}&\cite{chen20237}\setcounter{footnote}{1}\footnote{2}&\cite{dong202015}\setcounter{footnote}{1}\footnote{2}&SCOPE\cite{li2018scope}&DRISA\cite{li2017drisa}\footnote{3}&MAC-DO\_D &MAC-DO \\	
			
			\hline
			Tech&28nm&65nm&28nm&22nm&7nm&22nm&22nm&22nm&65nm \\
			\hline
			I/W Precision&INT8&16b fixedpoint&INT16,32&8b/8b&4b/4b&logic operation&logic operation&4b/4b&4b/4b \\
		    
		    \hline
		    Throughput& 40.4 TOPS&42 GOPS&5.58 TOPS&600 GOPS&372.4 GOPS&7.2 TOPS&1.68 TOPS&3.26 TOPS &6.4 GOPS    \\
		    
		    \hline
		    Workload & &CNN&CNN&CNN&CNN&CNN, RNN&CNN&CNN&CNN \\
		    
		    \hline
		    
		\end{tabular}
		
		\footnote[1] a Normalized to INT8, 
		\footnote[2] b Scaled to 65nm, assume energy and area $\propto$ Tech$^2$,
		\footnote[3] c re-evaluated to 8Gb capacity
  \vspace{-2em}
	\end{table*}
\end{scriptsize}

In this section, we estimate MAC-DO architecture's performance when it is used with a real DRAM array size using DRAM technology in Table \ref{table11} (=MAC-DO\_D). The array size is scaled to a typical DRAM MAT size (256$\times$512 MAC-DO cells, or 512$\times$512 1T1C DRAM cells \cite{zhang2014half}). Our estimation is based on the average power breakdown of C3 layer (Figure \ref{figure14}-(b)) and it scaled to 22nm DRAM technology. Then, we 
assume the average power is linear to the number of circuit blocks because most power dissipation is due to the dynamic power consumption, which is basically proportional to the size of parasitic capacitors. Overheads for controlling complicated row and column peripheral circuits are amortized over more number of MAC-DO cells, so The estimated performance of MAC-DO\_D demonstrates a 2.33$\times$ improvement in energy efficiency compared to the 16$\times$16 test circuit (Table \ref{table8}) and achieves a throughput of 3.26 TOPS, surpassing the performance of the 16$\times$16 array in CMOS technology.

\subsection{Area Estimation of MAC-DO}
\begin{figure}[h]
	\centering
 \vspace{-0.8em}
	\includegraphics[width=0.84\columnwidth]{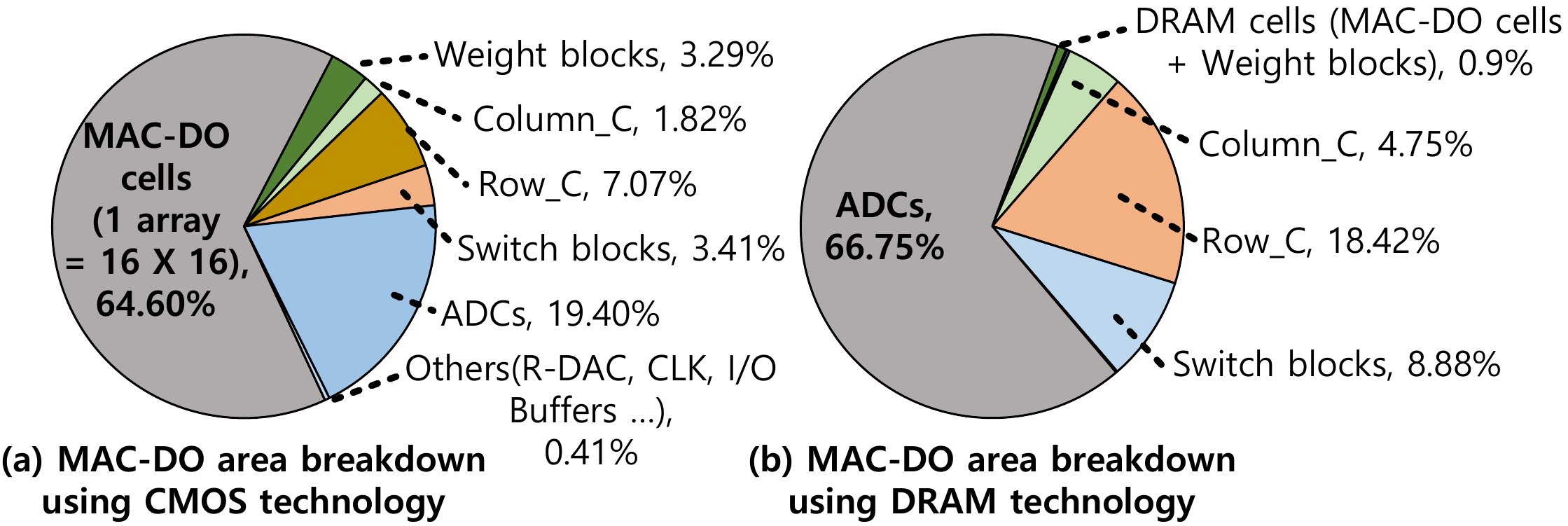}
	\caption{Area breakdown of MAC-DO architecture}
	\label{figure16}
\end{figure}
The area of each block in the MAC-DO test circuit has been estimated based on the transistor-level circuit design and layout using Cadence Virtuoso Layout Suite \cite{cadence2}. The total area is estimated 0.096mm$^2$ and Figure \ref{figure16}-(a) shows the area breakdown. Currently, most of the area is for the 16$\times$16 array (64.6\%), mainly due to large areas for cell capacitors in CMOS logic technology. The area for the ADCs makes up 19.40\% of the total area. The ADC overheads can be minimized by sharing one ADC with multiple BLs through multiplexers at the expense of longer readout latency. The area for the weight blocks accounts for 3.29\% of the total area, which is also dominated by the tail capacitors. The area of switch blocks accounts for 3.41\% and the area of row controller is 7.07\% of the total area because of a lot of switches in row periphery. The overall area of the MAC-DO architecture can be further decreased by using DRAM technology because a DRAM cell capacitor is designed much denser than a capacitor in CMOS logic process. In addition, the size of an access transistor can be scaled down to an actual access transistor size in DRAM. Given common centroid layout, the mismatch effect can be further minimized as more cells are activated at the same time. So, minimizing mismatch is boiled down to selecting how many cells regardless of cell size, but it would lower the throughput. Also, the switch transistors in the row controller can be much smaller by using normally sized access transistors ($\sim$10$\times$ smaller even with common centroid cell mapping) in the array because smaller access transistors require less driving force for the switches. Figure \ref{figure16}-(b) provides an area breakdown of MAC-DO\_D. The total area of MAC-DO\_D (1 mat size) is 0.123mm$^2$, with each DRAM cell occupying 0.0036 \SI{}{\micro\meter}$^2$. Therefore, approximately 26 mats are required to be nearly equivalent to the area of 1 DRAM bank size. Although MAC-DO\_D introduces some area overhead to the DRAM mat, its exceptional performance, even with a small portion of banks or mats, outweighs the overhead when compared to other DRAM-based accelerators (Section \uppercase\expandafter{\romannumeral6}-E, G, H).

\subsection{System Level Performance Analysis and Comparison}

\begin{figure}[!t]
	\centering
	\includegraphics[width=0.76\columnwidth]{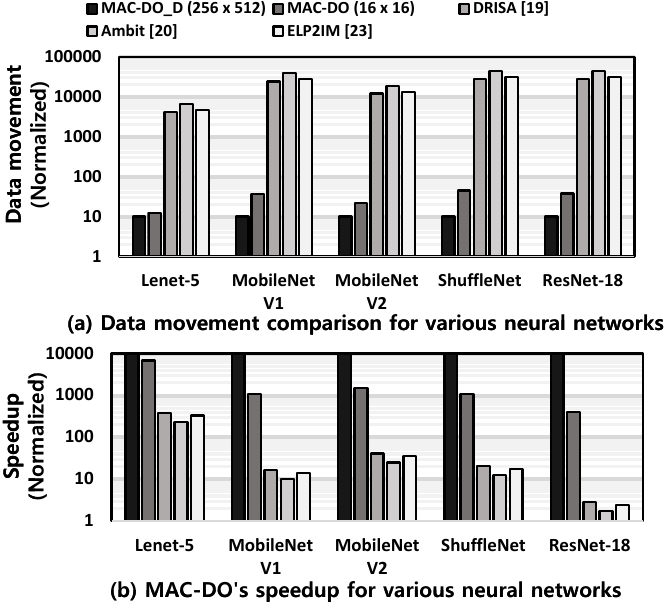}
	\caption{Data movement and speed comparisons with other DRAM-based in-situ accelerators for various neural networks}
 
\label{figure35}
\end{figure}

Figure \ref{figure35} compares the system performance of MAC-DO with previous DRAM-based in-situ accelerators (DRISA \cite{li2017drisa}, Ambit \cite{seshadri2017ambit}, ELP2IM \cite{xin2020elp2im}). The 16$\times$16 MAC-DO array shows \>300$\times$ data movement reduction compared with other accelerators as shown in Figure \ref{figure35}-(a). This is because all three types of data are efficiently reused for the entire matrix multiplication cycles and a single-cycle MAC operation reduces overall data access. Figure \ref{figure35}-(b) shows that the 16$\times$16 MAC-DO array can accelerate CNN layers \>17.9$\times$ faster than other accelerators thanks to its high throughput and a single-cycle MAC operation within a MAC-DO cell. In addition, as the MAC-DO array size increases, the system performances are improved thanks to higher data reusability and throughput.

\begin{figure}[!t]
	\centering
 \vspace{-1em}
	\includegraphics[width=\columnwidth]{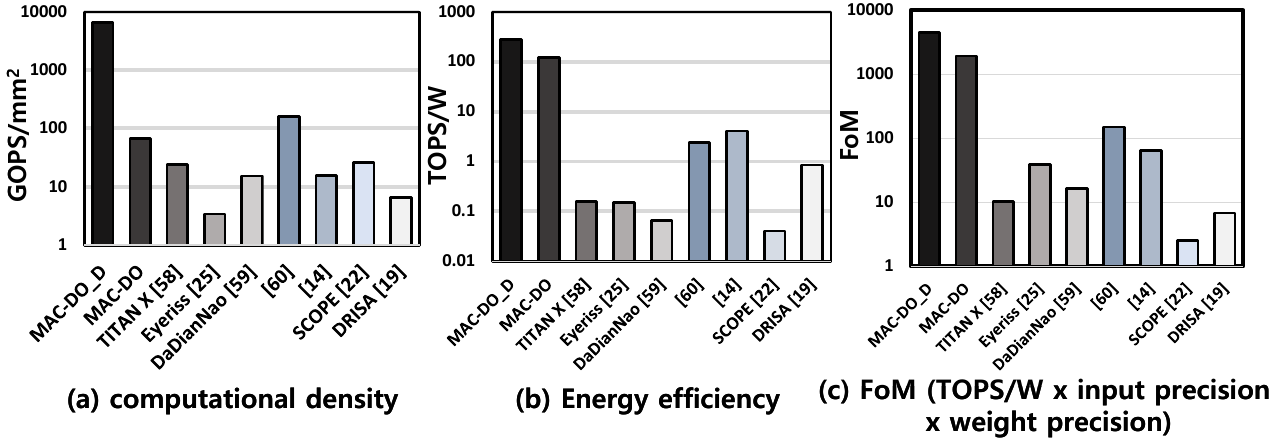}
	\caption{Performance comparison with other works}
	\label{figure19}
\end{figure}

\subsection{Performance Comparison with Other Works}

Figure \ref{figure19} compares the results with DRAM-based in-situ accelerators, a common GPU and other accelerators \cite{TITAN,chen2016eyeriss,chen2014dadiannao,chen20237,dong202015,li2018scope,li2017drisa}, and Table \ref{table6} shows their baseline descriptions for the comparison. Since a MAC-DO array size is 16$\times$16, the test chip has a low throughput, but in a real chip, it can be easily scaled to several TOPS as shown in Table \ref{table8}. Figure \ref{figure19}-(a) compares computational density (GOPS/mm{$^2$}) with other works. MAC-DO shows 2.55$\times$ computational density improvement over a recent DRAM-based in-situ accelerator \cite{li2018scope} thanks to high array utilization. In addition, the computational density will be further increased when MAC-DO is designed using DRAM technology. Figure \ref{figure19}-(b) shows that MAC-DO is at least minimum 29.7$\times$ more energy efficient than other compared works. Figure \ref{figure19}-(c) compares the FoM (energy efficiency(TOPS/W)$\times$input precision$\times$weight precision), and MAC-DO marks the best with $>$12.73$\times$ difference with previous works.

\section{Conclusion}
To overcome challenges of previous DRAM-based in-situ accelerators, this paper presents MAC-DO that performs analog multi-bit precision MAC operations directly using the 1T1C DRAM array and can achieve 100\% array utilization. The MAC operation happens inside each 2T2C MAC-DO cell and does not involve complex digital circuits, so it is well suited to DRAM technologies. The output stationary data flow allows efficient data reuse for all types of data. With the benefits, MAC-DO shows $>$2.55$\times$ computational density and $>$29.7$\times$ energy efficiency over all compared accelerators.


\bibliographystyle{IEEEtran}
\bibliography{refs}

\begin{thebibliography}{10}
\providecommand{\url}[1]{#1}
\csname url@samestyle\endcsname
\providecommand{\newblock}{\relax}
\providecommand{\bibinfo}[2]{#2}
\providecommand{\BIBentrySTDinterwordspacing}{\spaceskip=0pt\relax}
\providecommand{\BIBentryALTinterwordstretchfactor}{4}
\providecommand{\BIBentryALTinterwordspacing}{\spaceskip=\fontdimen2\font plus
\BIBentryALTinterwordstretchfactor\fontdimen3\font minus \fontdimen4\font\relax}
\providecommand{\BIBforeignlanguage}[2]{{%
\expandafter\ifx\csname l@#1\endcsname\relax
\typeout{** WARNING: IEEEtran.bst: No hyphenation pattern has been}%
\typeout{** loaded for the language `#1'. Using the pattern for}%
\typeout{** the default language instead.}%
\else
\language=\csname l@#1\endcsname
\fi
#2}}
\providecommand{\BIBdecl}{\relax}
\BIBdecl

\bibitem{shi2016edge}
W.~Shi, J.~Cao, Q.~Zhang, Y.~Li, and L.~Xu, ``Edge computing: Vision and challenges,'' \emph{IEEE internet of things journal}, vol.~3, no.~5, pp. 637--646, 2016.

\bibitem{satyanarayanan2017emergence}
M.~Satyanarayanan, ``The emergence of edge computing,'' \emph{Computer}, vol.~50, no.~1, pp. 30--39, 2017.

\bibitem{han2016eie}
S.~Han, X.~Liu, H.~Mao, J.~Pu, A.~Pedram, M.~A. Horowitz, and W.~J. Dally, ``Eie: Efficient inference engine on compressed deep neural network,'' \emph{ACM SIGARCH Computer Architecture News}, vol.~44, no.~3, pp. 243--254, 2016.

\bibitem{villa2014scaling}
O.~Villa, D.~R. Johnson, M.~Oconnor, E.~Bolotin, D.~Nellans, J.~Luitjens, N.~Sakharnykh, P.~Wang, P.~Micikevicius, A.~Scudiero \emph{et~al.}, ``Scaling the power wall: a path to exascale,'' in \emph{SC'14: Proceedings of the International Conference for High Performance Computing, Networking, Storage and Analysis}.\hskip 1em plus 0.5em minus 0.4em\relax IEEE, 2014, pp. 830--841.

\bibitem{collange2009power}
S.~Collange, D.~Defour, and A.~Tisserand, ``Power consumption of gpus from a software perspective,'' in \emph{International Conference on Computational Science}.\hskip 1em plus 0.5em minus 0.4em\relax Springer, 2009, pp. 914--923.

\bibitem{ahn2015scalable}
J.~Ahn, S.~Hong, S.~Yoo, O.~Mutlu, and K.~Choi, ``A scalable processing-in-memory accelerator for parallel graph processing,'' in \emph{Proceedings of the 42nd Annual International Symposium on Computer Architecture}, 2015, pp. 105--117.

\bibitem{xie2021spacea}
X.~Xie, Z.~Liang, P.~Gu, A.~Basak, L.~Deng, L.~Liang, X.~Hu, and Y.~Xie, ``Spacea: Sparse matrix vector multiplication on processing-in-memory accelerator,'' in \emph{2021 IEEE International Symposium on High-Performance Computer Architecture (HPCA)}.\hskip 1em plus 0.5em minus 0.4em\relax IEEE, 2021, pp. 570--583.

\bibitem{bojnordi2016memristive}
M.~N. Bojnordi and E.~Ipek, ``Memristive boltzmann machine: A hardware accelerator for combinatorial optimization and deep learning,'' in \emph{2016 IEEE International Symposium on High Performance Computer Architecture (HPCA)}.\hskip 1em plus 0.5em minus 0.4em\relax IEEE, 2016, pp. 1--13.

\bibitem{kim2018image}
Y.~Kim, M.~Imani, and T.~S. Rosing, ``Image recognition accelerator design using in-memory processing,'' \emph{IEEE Micro}, vol.~39, no.~1, pp. 17--23, 2018.

\bibitem{gu2020ipim}
P.~Gu, X.~Xie, Y.~Ding, G.~Chen, W.~Zhang, D.~Niu, and Y.~Xie, ``ipim: Programmable in-memory image processing accelerator using near-bank architecture,'' in \emph{2020 ACM/IEEE 47th Annual International Symposium on Computer Architecture (ISCA)}.\hskip 1em plus 0.5em minus 0.4em\relax IEEE, 2020, pp. 804--817.

\bibitem{imani2019floatpim}
M.~Imani, S.~Gupta, Y.~Kim, and T.~Rosing, ``Floatpim: In-memory acceleration of deep neural network training with high precision,'' in \emph{2019 ACM/IEEE 46th Annual International Symposium on Computer Architecture (ISCA)}.\hskip 1em plus 0.5em minus 0.4em\relax IEEE, 2019, pp. 802--815.

\bibitem{srivastava2018promise}
P.~Srivastava, M.~Kang, S.~K. Gonugondla, S.~Lim, J.~Choi, V.~Adve, N.~S. Kim, and N.~Shanbhag, ``Promise: An end-to-end design of a programmable mixed-signal accelerator for machine-learning algorithms,'' in \emph{2018 ACM/IEEE 45th Annual International Symposium on Computer Architecture (ISCA)}.\hskip 1em plus 0.5em minus 0.4em\relax IEEE, 2018, pp. 43--56.

\bibitem{biswas2018conv}
A.~Biswas and A.~P. Chandrakasan, ``Conv-ram: An energy-efficient sram with embedded convolution computation for low-power cnn-based machine learning applications,'' in \emph{2018 IEEE International Solid-State Circuits Conference-(ISSCC)}.\hskip 1em plus 0.5em minus 0.4em\relax IEEE, 2018, pp. 488--490.

\bibitem{dong202015}
Q.~Dong, M.~E. Sinangil, B.~Erbagci, D.~Sun, W.-S. Khwa, H.-J. Liao, Y.~Wang, and J.~Chang, ``15.3 a 351tops/w and 372.4 gops compute-in-memory sram macro in 7nm finfet cmos for machine-learning applications,'' in \emph{2020 IEEE International Solid-State Circuits Conference-(ISSCC)}.\hskip 1em plus 0.5em minus 0.4em\relax IEEE, 2020, pp. 242--244.

\bibitem{TPUv2v39351692}
T.~Norrie, N.~Patil, D.~H. Yoon, G.~Kurian, S.~Li, J.~Laudon, C.~Young, N.~Jouppi, and D.~Patterson, ``The design process for google's training chips: Tpuv2 and tpuv3,'' \emph{IEEE Micro}, vol.~41, no.~2, pp. 56--63, 2021.

\bibitem{horowitz20141}
M.~Horowitz, ``1.1 computing's energy problem (and what we can do about it),'' in \emph{2014 IEEE International Solid-State Circuits Conference Digest of Technical Papers (ISSCC)}.\hskip 1em plus 0.5em minus 0.4em\relax IEEE, 2014, pp. 10--14.

\bibitem{shin2018mcdram}
H.~Shin, D.~Kim, E.~Park, S.~Park, Y.~Park, and S.~Yoo, ``Mcdram: Low latency and energy-efficient matrix computations in dram,'' \emph{IEEE Transactions on Computer-Aided Design of Integrated Circuits and Systems}, vol.~37, no.~11, pp. 2613--2622, 2018.

\bibitem{lee20221ynm}
S.~Lee, K.~Kim, S.~Oh, J.~Park, G.~Hong, D.~Ka, K.~Hwang, J.~Park, K.~Kang, J.~Kim \emph{et~al.}, ``A 1ynm 1.25 v 8gb, 16gb/s/pin gddr6-based accelerator-in-memory supporting 1tflops mac operation and various activation functions for deep-learning applications,'' in \emph{2022 IEEE International Solid-State Circuits Conference (ISSCC)}, vol.~65.\hskip 1em plus 0.5em minus 0.4em\relax IEEE, 2022, pp. 1--3.

\bibitem{li2017drisa}
S.~Li, D.~Niu, K.~T. Malladi, H.~Zheng, B.~Brennan, and Y.~Xie, ``Drisa: A dram-based reconfigurable in-situ accelerator,'' in \emph{2017 50th Annual IEEE/ACM International Symposium on Microarchitecture (MICRO)}.\hskip 1em plus 0.5em minus 0.4em\relax IEEE, 2017, pp. 288--301.

\bibitem{seshadri2017ambit}
V.~Seshadri, D.~Lee, T.~Mullins, H.~Hassan, A.~Boroumand, J.~Kim, M.~A. Kozuch, O.~Mutlu, P.~B. Gibbons, and T.~C. Mowry, ``Ambit: In-memory accelerator for bulk bitwise operations using commodity dram technology,'' in \emph{2017 50th Annual IEEE/ACM International Symposium on Microarchitecture (MICRO)}.\hskip 1em plus 0.5em minus 0.4em\relax IEEE, 2017, pp. 273--287.

\bibitem{angizi2019redram}
S.~Angizi and D.~Fan, ``Redram: A reconfigurable processing-in-dram platform for accelerating bulk bit-wise operations,'' in \emph{2019 IEEE/ACM International Conference on Computer-Aided Design (ICCAD)}.\hskip 1em plus 0.5em minus 0.4em\relax IEEE, 2019, pp. 1--8.

\bibitem{li2018scope}
S.~Li, A.~O. Glova, X.~Hu, P.~Gu, D.~Niu, K.~T. Malladi, H.~Zheng, B.~Brennan, and Y.~Xie, ``Scope: A stochastic computing engine for dram-based in-situ accelerator,'' in \emph{2018 51st Annual IEEE/ACM International Symposium on Microarchitecture (MICRO)}.\hskip 1em plus 0.5em minus 0.4em\relax IEEE, 2018, pp. 696--709.

\bibitem{xin2020elp2im}
X.~Xin, Y.~Zhang, and J.~Yang, ``Elp2im: Efficient and low power bitwise operation processing in dram,'' in \emph{2020 IEEE International Symposium on High Performance Computer Architecture (HPCA)}.\hskip 1em plus 0.5em minus 0.4em\relax IEEE, 2020, pp. 303--314.

\bibitem{kim1999assessing}
Y.-B. Kim and T.~W. Chen, ``Assessing merged dram/logic technology,'' \emph{Integration}, vol.~27, no.~2, pp. 179--194, 1999.

\bibitem{chen2016eyeriss}
Y.-H. Chen, J.~Emer, and V.~Sze, ``Eyeriss: A spatial architecture for energy-efficient dataflow for convolutional neural networks,'' \emph{ACM SIGARCH Computer Architecture News}, vol.~44, no.~3, pp. 367--379, 2016.

\bibitem{razavi2013charge}
B.~Razavi, ``Charge steering: A low-power design paradigm,'' in \emph{Proceedings of the IEEE 2013 Custom Integrated Circuits Conference}.\hskip 1em plus 0.5em minus 0.4em\relax IEEE, 2013, pp. 1--8.

\bibitem{moons201714}
B.~Moons, R.~Uytterhoeven, W.~Dehaene, and M.~Verhelst, ``14.5 envision: A 0.26-to-10tops/w subword-parallel dynamic-voltage-accuracy-frequency-scalable convolutional neural network processor in 28nm fdsoi,'' in \emph{2017 IEEE International Solid-State Circuits Conference (ISSCC)}.\hskip 1em plus 0.5em minus 0.4em\relax IEEE, 2017, pp. 246--247.

\bibitem{du2015shidiannao}
Z.~Du, R.~Fasthuber, T.~Chen, P.~Ienne, L.~Li, T.~Luo, X.~Feng, Y.~Chen, and O.~Temam, ``Shidiannao: Shifting vision processing closer to the sensor,'' in \emph{Proceedings of the 42nd Annual International Symposium on Computer Architecture}, 2015, pp. 92--104.

\bibitem{reagen2021cheetah}
B.~Reagen, W.-S. Choi, Y.~Ko, V.~T. Lee, H.-H.~S. Lee, G.-Y. Wei, and D.~Brooks, ``Cheetah: Optimizing and accelerating homomorphic encryption for private inference,'' in \emph{2021 IEEE International Symposium on High-Performance Computer Architecture (HPCA)}.\hskip 1em plus 0.5em minus 0.4em\relax IEEE, 2021, pp. 26--39.

\bibitem{venkataramani2021rapid}
S.~Venkataramani, V.~Srinivasan, W.~Wang, S.~Sen, J.~Zhang, A.~Agrawal, M.~Kar, S.~Jain, A.~Mannari, H.~Tran \emph{et~al.}, ``Rapid: Ai accelerator for ultra-low precision training and inference,'' in \emph{2021 ACM/IEEE 48th Annual International Symposium on Computer Architecture (ISCA)}.\hskip 1em plus 0.5em minus 0.4em\relax IEEE, 2021, pp. 153--166.

\bibitem{vasudevan2017parallel}
A.~Vasudevan, A.~Anderson, and D.~Gregg, ``Parallel multi channel convolution using general matrix multiplication,'' in \emph{2017 IEEE 28th international conference on application-specific systems, architectures and processors (ASAP)}.\hskip 1em plus 0.5em minus 0.4em\relax IEEE, 2017, pp. 19--24.

\bibitem{ofir2022smm}
A.~Ofir and G.~Ben-Artzi, ``Smm-conv: Scalar matrix multiplication with zero packing for accelerated convolution,'' in \emph{Proceedings of the IEEE/CVF Conference on Computer Vision and Pattern Recognition}, 2022, pp. 3067--3075.

\bibitem{pal2018outerspace}
S.~Pal, J.~Beaumont, D.-H. Park, A.~Amarnath, S.~Feng, C.~Chakrabarti, H.-S. Kim, D.~Blaauw, T.~Mudge, and R.~Dreslinski, ``Outerspace: An outer product based sparse matrix multiplication accelerator,'' in \emph{2018 IEEE International Symposium on High Performance Computer Architecture (HPCA)}.\hskip 1em plus 0.5em minus 0.4em\relax IEEE, 2018, pp. 724--736.

\bibitem{jung201325}
J.~W. Jung and B.~Razavi, ``A 25-gb/s 5-mw cmos cdr/deserializer,'' \emph{IEEE Journal of Solid-State Circuits}, vol.~48, no.~3, pp. 684--697, 2013.

\bibitem{chiang201410}
S.-H.~W. Chiang, H.~Sun, and B.~Razavi, ``A 10-bit 800-mhz 19-mw cmos adc,'' \emph{IEEE Journal of Solid-State Circuits}, vol.~49, no.~4, pp. 935--949, 2014.

\bibitem{deng2018dracc}
Q.~Deng, L.~Jiang, Y.~Zhang, M.~Zhang, and J.~Yang, ``Dracc: a dram based accelerator for accurate cnn inference,'' in \emph{Proceedings of the 55th Annual Design Automation Conference}, 2018, pp. 1--6.

\bibitem{shin2019pvt}
H.~Shin, J.~Sim, D.~Lee, and L.-S. Kim, ``A pvt-robust customized 4t embedded dram cell array for accelerating binary neural networks,'' in \emph{2019 IEEE/ACM International Conference on Computer-Aided Design (ICCAD)}.\hskip 1em plus 0.5em minus 0.4em\relax IEEE, 2019, pp. 1--8.

\bibitem{roy2021pim}
S.~Roy, M.~Ali, and A.~Raghunathan, ``Pim-dram: Accelerating machine learning workloads using processing in commodity dram,'' \emph{IEEE Journal on Emerging and Selected Topics in Circuits and Systems}, vol.~11, no.~4, pp. 701--710, 2021.

\bibitem{bastos1996matching}
J.~Bastos, M.~Steyaert, B.~Graindourze, and W.~Sansen, ``Matching of mos transistors with different layout styles,'' in \emph{Proceedings of International Conference on Microelectronic Test Structures}.\hskip 1em plus 0.5em minus 0.4em\relax IEEE, 1996, pp. 17--18.

\bibitem{van20012}
H.~Van~der Ploeg, G.~Hoogzaad, H.~A. Termeer, M.~Vertregt, and R.~L. Roovers, ``A 2.5-v 12-b 54-msample/s 0.25-/spl mu/m cmos adc in 1-mm/sup 2/with mixed-signal chopping and calibration,'' \emph{IEEE Journal of Solid-State Circuits}, vol.~36, no.~12, pp. 1859--1867, 2001.

\bibitem{joksas2022nonideality}
D.~Joksas, E.~Wang, N.~Barmpatsalos, W.~H. Ng, A.~J. Kenyon, G.~A. Constantinides, and A.~Mehonic, ``Nonideality-aware training for accurate and robust low-power memristive neural networks,'' \emph{Advanced Science}, vol.~9, no.~17, p. 2105784, 2022.

\bibitem{paszke2017automatic}
A.~Paszke, S.~Gross, S.~Chintala, G.~Chanan, E.~Yang, Z.~DeVito, Z.~Lin, A.~Desmaison, L.~Antiga, and A.~Lerer, ``Automatic differentiation in pytorch,'' 2017.

\bibitem{jesd}
``Ddr4 sdram specification.''\hskip 1em plus 0.5em minus 0.4em\relax http://www.softnology.biz/pdf/JESD79-4B.pdf.

\bibitem{choi2019optimized}
S.~Choi, J.~Shin, Y.~Choi, and L.-S. Kim, ``An optimized design technique of low-bit neural network training for personalization on iot devices,'' in \emph{Proceedings of the 56th Annual Design Automation Conference 2019}, 2019, pp. 1--6.

\bibitem{mishra2017apprentice}
A.~Mishra and D.~Marr, ``Apprentice: Using knowledge distillation techniques to improve low-precision network accuracy,'' \emph{arXiv preprint arXiv:1711.05852}, 2017.

\bibitem{choukroun2019low}
Y.~Choukroun, E.~Kravchik, F.~Yang, and P.~Kisilev, ``Low-bit quantization of neural networks for efficient inference,'' in \emph{2019 IEEE/CVF International Conference on Computer Vision Workshop (ICCVW)}.\hskip 1em plus 0.5em minus 0.4em\relax IEEE, 2019, pp. 3009--3018.

\bibitem{cadence}
``Cadence spectre circuit simulator.''\hskip 1em plus 0.5em minus 0.4em\relax www.cadence.com.

\bibitem{cadence1}
``Cadence virtuoso schematic editor.''\hskip 1em plus 0.5em minus 0.4em\relax www.cadence.com.

\bibitem{ADCcite}
B.~Murmann, ``Adc performance survey 1997-2021 (isscc \& vlsi symposium).''\hskip 1em plus 0.5em minus 0.4em\relax https://web.stanford.edu/~murmann/adcsurvey.html, 2021.

\bibitem{fitzpatrick1998analog}
D.~FitzPatrick and I.~Miller, \emph{Analog behavioral modeling with the Verilog-A language}.\hskip 1em plus 0.5em minus 0.4em\relax Springer Science \& Business Media, 1998.

\bibitem{mnistdataset}
Y.~LeCun, C.~Cortes, and C.~Burges, ``The mnist database of handwritten digits.''\hskip 1em plus 0.5em minus 0.4em\relax http://yann.lecun.com/exdb/mnist/.

\bibitem{lecun1998gradient}
Y.~LeCun, L.~Bottou, Y.~Bengio, and P.~Haffner, ``Gradient-based learning applied to document recognition,'' \emph{Proceedings of the IEEE}, vol.~86, no.~11, pp. 2278--2324, 1998.

\bibitem{jung2022crossbar}
S.~Jung, H.~Lee, S.~Myung, H.~Kim, S.~K. Yoon, S.-W. Kwon, Y.~Ju, M.~Kim, W.~Yi, S.~Han \emph{et~al.}, ``A crossbar array of magnetoresistive memory devices for in-memory computing,'' \emph{Nature}, vol. 601, no. 7892, pp. 211--216, 2022.

\bibitem{howard2017mobilenets}
A.~G. Howard, M.~Zhu, B.~Chen, D.~Kalenichenko, W.~Wang, T.~Weyand, M.~Andreetto, and H.~Adam, ``Mobilenets: Efficient convolutional neural networks for mobile vision applications,'' \emph{arXiv preprint arXiv:1704.04861}, 2017.

\bibitem{sandler2018mobilenetv2}
M.~Sandler, A.~Howard, M.~Zhu, A.~Zhmoginov, and L.-C. Chen, ``Mobilenetv2: Inverted residuals and linear bottlenecks,'' in \emph{Proceedings of the IEEE conference on computer vision and pattern recognition}, 2018, pp. 4510--4520.

\bibitem{zhang2018shufflenet}
X.~Zhang, X.~Zhou, M.~Lin, and J.~Sun, ``Shufflenet: An extremely efficient convolutional neural network for mobile devices,'' in \emph{Proceedings of the IEEE conference on computer vision and pattern recognition}, 2018, pp. 6848--6856.

\bibitem{he2016deep}
K.~He, X.~Zhang, S.~Ren, and J.~Sun, ``Deep residual learning for image recognition,'' in \emph{Proceedings of the IEEE conference on computer vision and pattern recognition}, 2016, pp. 770--778.

\bibitem{TITAN}
``Nvidea gpu.''\hskip 1em plus 0.5em minus 0.4em\relax http://www.nvidia.com, 2016.

\bibitem{chen2014dadiannao}
Y.~Chen, T.~Luo, S.~Liu, S.~Zhang, L.~He, J.~Wang, L.~Li, T.~Chen, Z.~Xu, N.~Sun \emph{et~al.}, ``Dadiannao: A machine-learning supercomputer,'' in \emph{2014 47th Annual IEEE/ACM International Symposium on Microarchitecture}.\hskip 1em plus 0.5em minus 0.4em\relax IEEE, 2014, pp. 609--622.

\bibitem{chen20237}
P.~Chen, M.~Wu, W.~Zhao, J.~Cui, Z.~Wang, Y.~Zhang, Q.~Wang, J.~Ru, L.~Shen, T.~Jia \emph{et~al.}, ``7.8 a 22nm delta-sigma computing-in-memory sram macro with near-zero-mean outputs and lsb-first adcs achieving 21.38 tops/w for 8b-mac edge ai processing,'' in \emph{2023 IEEE International Solid-State Circuits Conference (ISSCC)}.\hskip 1em plus 0.5em minus 0.4em\relax IEEE, 2023, pp. 140--142.

\bibitem{zhang2014half}
T.~Zhang, K.~Chen, C.~Xu, G.~Sun, T.~Wang, and Y.~Xie, ``Half-dram: A high-bandwidth and low-power dram architecture from the rethinking of fine-grained activation,'' in \emph{2014 ACM/IEEE 41st International Symposium on Computer Architecture (ISCA)}.\hskip 1em plus 0.5em minus 0.4em\relax IEEE, 2014, pp. 349--360.

\bibitem{cadence2}
``Cadence virtuoso layout suite.''\hskip 1em plus 0.5em minus 0.4em\relax www.cadence.com.

\end{thebibliography}

\end{document}